\documentclass[fleqn,usenatbib]{mnras}
\usepackage[T1]{fontenc}
\usepackage{ae,aecompl}
\usepackage{graphicx}
\usepackage{amsmath}
\usepackage{amssymb}
\usepackage{txfonts}

\title[Colour-mass diagrams]
{
Repainting the colour-mass diagrams by unearthing the green mountain:
dust-rich S0 galaxies in the colour-(galaxy stellar mass) diagram,  
and the 
colour-(black hole mass) relations for dust-poor versus dust-rich galaxies
}

\author[Graham]
{
Alister W.\ Graham$^{1}$\thanks{E-mail: AGraham@swin.edu.au}
\\
$^1$ Centre for Astrophysics and Supercomputing, Swinburne University of
Technology, Hawthorn, VIC 3122, Australia 
}

\date{Accepted XXX. Received YYY; in original form ZZZ}
\pubyear{2024}

\begin{document}
\label{firstpage}
\pagerange{\pageref{firstpage}--\pageref{lastpage}}
\maketitle

\begin{abstract}

Lenticular galaxies are notoriously misclassified as elliptical galaxies and,
as such, a (disc inclination)-dependent correction for dust is often not
applied to the magnitudes of dusty lenticular galaxies. This results in overly
red galaxy colours, impacting their distribution in the colour-magnitude
diagram.  It is revealed how this has led to an underpopulation of the `green
valley' by hiding a `green mountain' of massive dust-rich lenticular
galaxies---known to be built from gas-rich major mergers---within the `red
sequence' of colour-(stellar mass) diagrams.  Correcting for dust, a `green
mountain' appears at $M_{\rm *,gal}\sim10^{11}$~M$_\odot$, along with signs of
an extension to lower masses producing a `green range' or `green ridge' on the
green side of the `red sequence' and `blue cloud.'  The `red sequence' is
shown to be comprised of two components: a red plateau defined by elliptical
galaxies with a near-constant colour and by lower-mass dust-poor lenticular
galaxies, which are mostly a primordial population but may include
faded/transformed spiral galaxies.  The
presence of the
quasi-triangular-shaped galaxy evolution sequence, previously called the
`Triangal', is revealed in the galaxy colour-(stellar mass) diagram.  It
tracks the speciation of galaxies and their associated migration through the
diagram. The connection of the `Triangal' to previous galaxy morphology
sequences (Fork, Trident, Comb) is also shown herein. Finally, the colour-(black hole mass) diagram
is revisited, revealing how the dust correction generates a blue-green
sequence for the spiral {\em and} dust-rich lenticular galaxies that is offset
from a green-red sequence defined by the dust-poor lenticular and elliptical
galaxies.


\end{abstract}


\section{Introduction}

\begin{figure*}
\begin{center}
\includegraphics[trim=0.0cm 0cm 0.0cm 0cm, width=0.75\textwidth,
  angle=0]{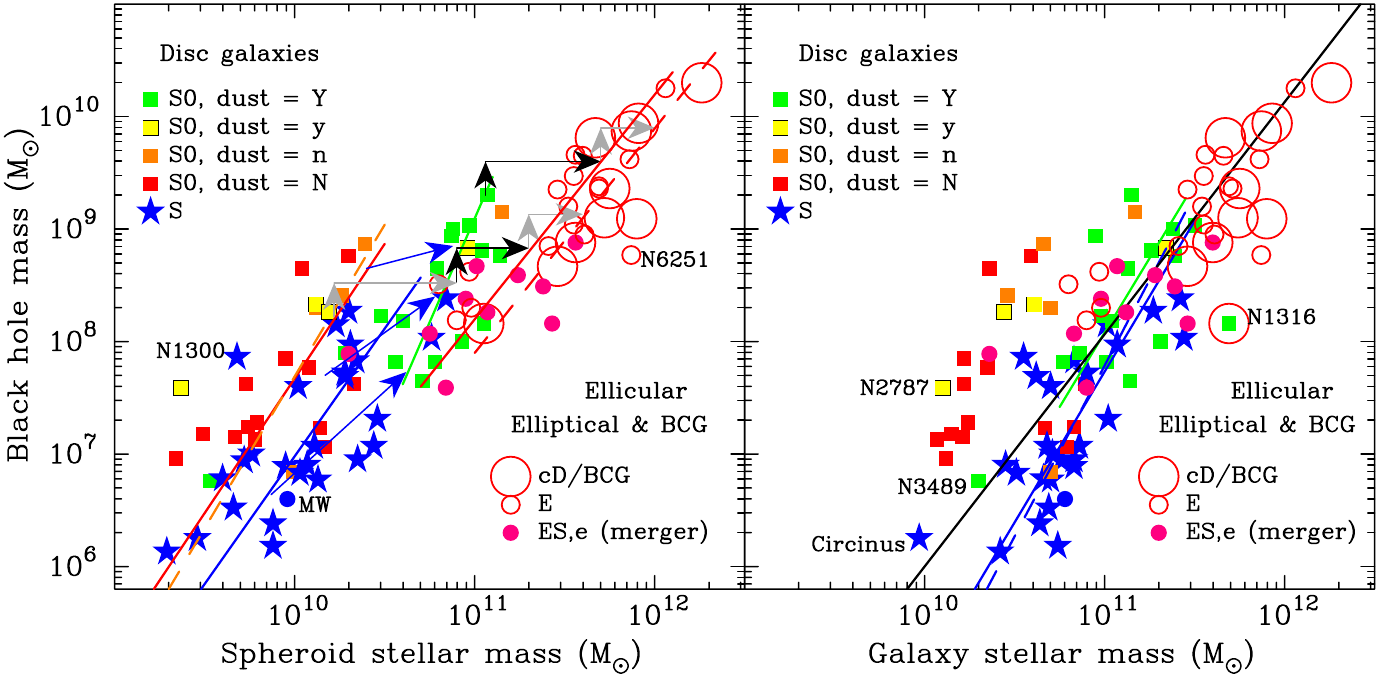}
\caption{Adaption of figures 1 and A4 from \citet{Graham-triangal}.  Left
  panel: From left to right, the red (and dashed orange) line pertains to the dust$=$(N)o S0
  galaxies (and when additionally including the dust$=$(n)ucleus-only S0 galaxies), while the
  other line colours are for the S (blue), S0 with dust$=$(Y)es-strong
  (green), and E plus ES,e (red) galaxies.  The dashed red line is
  representative of the BCG.
  The grey and black arrows denote growth from major (equal mass) dry
  mergers, while the blue arrows denote growth from major gas-rich mergers.  Right
  panel: The solid (and dashed) blue line corresponds to the S galaxies (after
  excluding Circinus). The green line pertains to S0 galaxies with dust$=$Y.
  (Arguably, the S and dust-rich S0 galaxies define a common sequence, and they
  have been combined in the CM$_{\rm bh}$D (Section~\ref{Sec_col_bh}); many of the
  dust-rich S0 galaxies have mild levels of star formation \citep{Graham-SFR}; 
  see also \citet{2005ApJ...619L.111Y}.)  The black line pertains to the
  cD/BCG, E, ES,e and dust$=$Y S0 galaxies and effectively represents the
  major-merger built galaxies.  Slopes are around 2 to 3, with equations and
  details given in \citet{Graham-triangal}.  }
\label{Fig0}
\end{center}
\end{figure*}

Galaxy speciation, i.e., the evolution of galaxies from one type to another,
tracks, among other things, the uni-directional growth
of their central black hole\footnote{For massive
black holes, 
`Hawking radiation' \citep{1974Natur.248...30H, 1975CMaPh..43..199H,
  1976PZETF..24...29Z}---first discussed by Vladimir Gribov and Yakov
Zel'dovich in 1972-1973 \citep{Anselm98, Dokshitzer98, Lipatov99,
  2016IJMPA..3145004A}---tends to be dramatically outpaced by black hole
accretions and mergers.} mass, $M_{\rm bh}$, and
bulge/spheroid stellar mass, $M_{\rm *,sph}$.  Galaxy-merger-induced jumps
from one type of galaxy to another was initially observed
\citep{2012ApJ...746..113G, 2013ApJ...764..151G, 2013ApJ...768...76S} 
in the $M_{\rm bh}$-$M_{\rm *,sph}$ diagram due to the appearance of 
merger-built galaxies with cores depleted of stars 
following a different trend to those without such central
deficits.\footnote{This galaxy divide at around $2\times10^{11}$
M$_{\odot}$ is discussed in \citet{2003AJ....125.2936G}, while the history of 
black hole scaling relations up to and including this separation of galaxy
type is extensively reviewed in \citet{2016ASSL..418..263G}.}
Others have subsequently confirmed this scenario, including 
\citet[][their Section~6.7]{2013ARA&A..51..511K} and \citet{2016ApJ...818...47S}. 
Today, a more complete picture of these transitions is emerging,
revealing transitions from spiral (S) to dust-rich
lenticular \citep[S0: e.g.,][]{1996ApJ...471..115B, 2008ApJ...685L..27R} and  from dust-rich S0 to
either ellicular (ES)\footnote{\citet{1966ApJ...146...28L} introduced the ES
notation to capture something of a halfway station between E and (massive) S0
galaxies.  The term `ellicular' was introduced/preferred
\citep{2016ApJ...831..132G} over `lentical' given that the ES galaxy sample seemed
more akin to E than S0 galaxies, with disc-to-total ratios generally less
than 15 per cent \citep{Graham:Sahu:22b}.}  or elliptical (E) galaxies \citep{1993ApJ...409..548H},
as described in  \citet{Graham-triangal} and referred to 
as `punctuated equilibrium' by \citet{Graham:Sahu:22a}.  For convenience,
these galaxy 
transitions are illustrated in Fig.~\ref{Fig0}, emerging from the
(galaxy morphology)-aware $M_{\rm bh}$-$M_{\rm *,sph}$ diagram shown there. 
All common types of galaxies are now considered, and previous reasons for
excluding significant numbers of galaxies are eliminated. These reasons
included not following a nearly linear $M_{\rm bh}$-$M_{\rm *,sph}$
relationship due to an `over-massive black hole' monster, having a
suspected pseudobulge (regardless of the presence of a classical bulge) or
being a merger
remnant, and so on \citep{2013ARA&A..51..511K}.
A more inclusive and unified scheme is rewriting our 
understanding of the coevolution of galaxies and their massive black holes.

\begin{figure*}
\begin{center}
\includegraphics[trim=0.0cm 0cm 0.0cm 0cm, width=0.7\textwidth, angle=0]{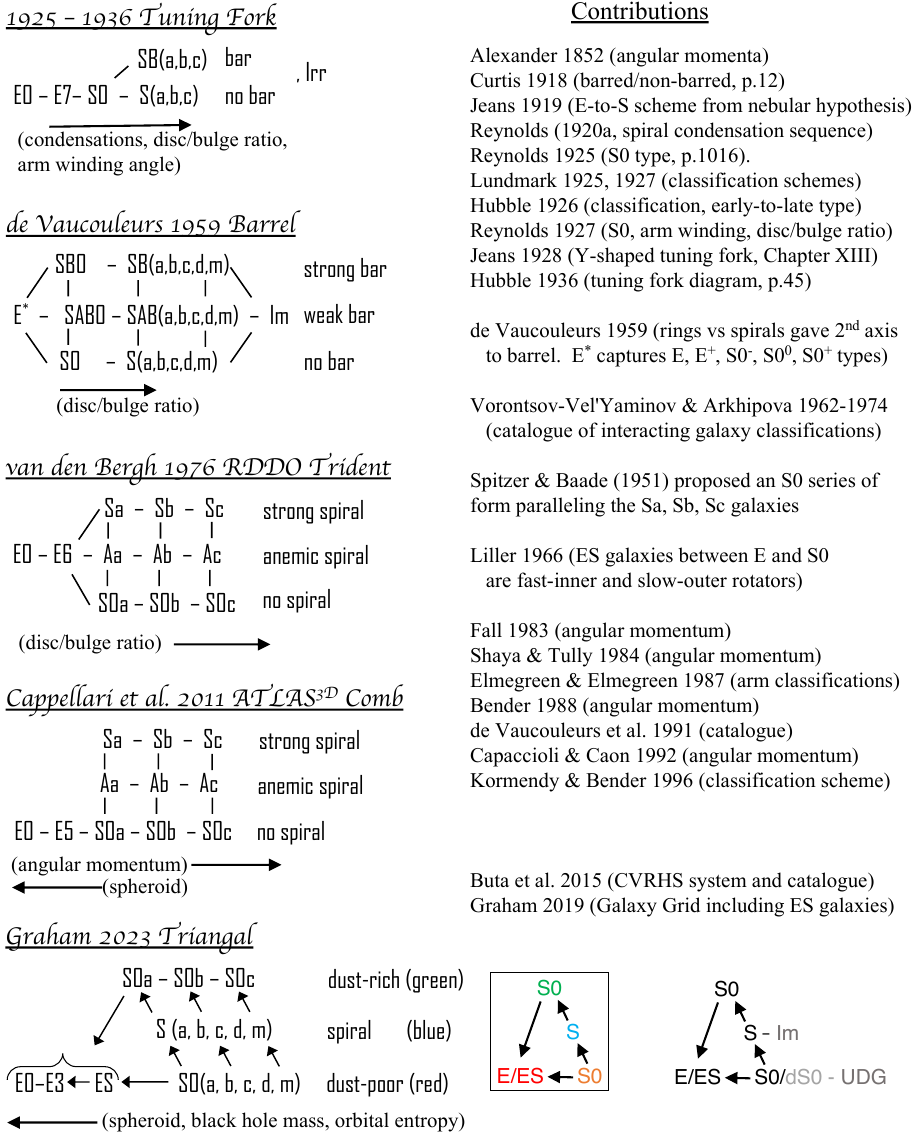}
\caption{The accretion/merger-driven transitions to new morphological types
  --- shown by 
  the `Triangal' \citep{Graham-triangal} --- are evident in Fig.~\ref{Fig0} and track increases in
 a galaxy's spheroid (aka bulge) and black hole mass.  The transitions reveal the dust-rich S0 galaxies built
  from wet major mergers as a different population to the lower-mass dust-poor
  S0 galaxies, perhaps better regarded as `first generation' galaxies rather
  than just `primordials', thereby encapsulating delayed creation at later
  epochs. The transitions also show S galaxies (and Magellanic-like Irregular
  galaxies: Im) developed from accretions and
  gravitational perturbations of previously spiral-less disc galaxies. Major
  mergers of S0 galaxies build the ES and E galaxies. 
The extended schema, including dwarf S0 galaxies and ultra-diffuse galaxies
(UDGs) in the lower right, shall be addressed in a forthcoming
paper.   Notes:
  RDDO = Revised David Dunlap Observatory. 
  CVRHS system = Comprehensive de Vaucouleurs revised Hubble-Sandage (CVRHS) system. 
  The location of disc galaxies with (i) weak vs. strong spiral
  arms, (ii) weak vs. strong bars, and (iii) ring/spiral variety
  \citep{1959HDP....53..275D} 
  in this accretion/merger-driven evolutionary schema will be explored in forthcoming work. 
  References: \citep{1852AJ......2...95A, 1918PLicO..13....9C,
    1919pcsd.book.....J, 1920MNRAS..80..746R, 1925MNRAS..85..865L,
    1925MNRAS..85.1014R, 1926ApJ....64..321H, 1927UGC..........1L,
    1927Obs....50..185R, 1928asco.book.....J, 1936rene.book.....H,
    1959HDP....53..275D, 1962MCG...C01....0V, 1966ApJ...146...28L, 1974TrSht..46....1V,
    1976ApJ...206..883V, 1983IAUS..100..391F, 1984ApJ...281...56S,
    1987ApJ...314....3E, 1988A&A...193L...7B, 1991rc3..book.....D,
    1992ASSL..178...99C, 1996ApJ...464L.119K, 2011MNRAS.416.1680C,
    2015ApJS..217...32B, 2019MNRAS.487.4995G, Graham-triangal}.  
}
\label{Fig_schemas}
\end{center}
\end{figure*}

The ubiquitous nature of massive black holes in regular galaxies
was realised three decades ago
\citep[e.g.,][]{1993MNRAS.263..168H} and has been supported by a continual stream of
observations \citep[e.g.,][]{1995Natur.375..659T, 1995Natur.373..127M,
  1996ApJ...472..153G, 1997MNRAS.291..219G, 1998ApJ...509..678G,
  2002Natur.419..694S, 2019ApJ...875L...4E}.
As black hole masses increase, so does their entropy \citep{1973PhRvD...7.2333B}. 
A spheroid's stellar mass reflects its galaxy's orbital entropy\footnote{This
remark excludes the entropy of suspected dark matter halos of
non-baryonic matter \citep{1981ApJ...246..557O, 1999PhRvL..82.2644F},
including gravitinos \citep{1941PhRv...60...61R, 1977PhRvD..15..996G,
  1982PhRvL..48.1636B, 1982PhRvL..48..223P} and 
potential primordial low- 
and intermediate-mass black holes \citep{1971MNRAS.152...75H,
  1974MNRAS.168..399C, 2020ARNPS..70..355C}
missed by gravitational lensing surveys \citep[e.g.,][]{2000ApJ...542..281A,
  2007A&A...469..387T}
because they are either too small or too few, respectively.}, 
a proxy for  
disorder and chaos over ordered disc orbits.  It is a measure of
the dispersion of the orbital energy into more configurations, better filling
the spherical volume or halo enclosing a galaxy.
This greater appreciation of spheroid-building mergers and the
coevolution of their central black hole helped solve a century-long
mystery of galaxy evolution, giving rise to a reworked `Tuning Fork' diagram
\citep{1928asco.book.....J, 1936rene.book.....H} 
referred to as the `Triangal' and shown in Fig.~\ref{Fig_schemas}.

Among its successes, the `Triangal' has helped better link 
the Milky Way with the extragalactic population \citep{Graham-triangal}.
Building on the Tuning Fork, the `Triangal' strikes a new note by including
evolutionary pathways due to acquisitions and mergers and involving a 
more extensive array of galaxy types, including ellicular galaxies and two or three
varieties of S0 galaxies.  There are the dust-rich S0 galaxies built from
gas-rich, aka `wet', 
mergers involving S galaxies (already polluted/fertilised with dust) 
\citep[e.g.,][]{1996ApJ...471..115B, 2017MNRAS.471.1892B, 2022MNRAS.513..389R},
and the dust-poor S0 galaxies comprised of (some) 
faded S galaxies \citep{1972ApJ...176....1G, 2007A&A...470..173B} 
and (mostly) primordial S0 galaxies, which can have old, 
metal-poor stellar populations \citep[e.g.,][]{2004AJ....127.1502R,
  2004AJ....127.3213M, 2010MNRAS.405..800P} and low bulge-to-total ($B/T$)
stellar mass ratios. 
The latter population need not all be old, though, and they might be better regarded as the first
generation/incarnation of a galaxy, in which gas cooled sufficiently to contract into a disc (or ring) to allow
star formation.  They need never have possessed a spiral pattern.
Those S0 galaxies that formed in the young Universe and did not
transform their morphology could be thought of as today's 
`ancients' or `elders'
\citep{2006AJ....132.2432L, 2013MSAIS..25...93S}.  Those which formed
at later times or experienced replenishment/rejuvenation
\citep{2015Galax...3..192M, 2015A&A...575A..16M, 2022MNRAS.513..389R} of their stellar ranks 
will appear younger, while local gas-dominated `dark galaxies' may represent
current-epoch counterparts \citep[e.g.,][]{1988ApJ...332L..33C,
  2001AJ....121.1461K}. 

The `Triangal' not only encapsulates the joint development of galaxies and
black holes and has breathed life (clear evolutionary pathways) 
into past sequences of galaxy morphology (Fig.~\ref{Fig_schemas}), but also 
applies to interpreting the (star formation rate: SFR)-(stellar mass)
diagram \citep{Graham-SFR}.  The current paper provides two additional applications
of the `Triangal'.  First, its presence and operation are revealed  in the galaxy
colour-(stellar mass) diagram
\citep[CMD: e.g.,][]{1959PASP...71..106B, 1961ApJS....5..233D, 1977ApJ...216..214V,
  1978ApJ...225..742S, 1997A&A...320...41K, 2007ApJS..173..342M,
  2007ApJS..173..315S, 2007ApJS..173..293W}.
It must be noted that this paper benefits from
countless prior works, and a sincere effort to acknowledge some of this
relevant and interesting history is presented in Section~\ref{Sec_Hist}
before the data for the present investigation is described in Section~\ref{Sec_data}.

The CMD is often touted as a tool for the wholesale study of the extragalactic
constituents of the cosmos.  
Nowadays, the CMD is typically considered to have three main features: a `red
sequence', `green valley' \citep[e.g.,][and references
  therein]{2007ApJS..173..342M, 2014SerAJ.189....1S} or `green plain'
\citep{2022A&A...666A.170Q}, and `blue cloud'. 
In Section~\ref{Sec_Results}, 
with attention to a greater range of galaxy types and using dust corrections
for both late-type galaxies (LTGs) and early-type galaxies (ETGs) 
with dust-rich discs, the `green valley/plain' is somewhat 
replaced by a `green mountain' at $M_{\rm *,gal}\sim10^{11}$~M$_\odot$ 
\citep{2005ApJ...619L.111Y, 2018MNRAS.481.1183E, Graham-SFR},
with an extension to lower masses forming what
might be termed a `green range' or `green ridge'. This extension was previously highlighted by
\citet{1992AJ....104.1039S} as due to a population later referred to as blue ETGs
\citep[e.g.,][]{2006AJ....132.2432L, 2007ApJ...657L..85D, 2009AJ....138..579K}. 
In addition, a `red plateau' or flat `dead sequence'\footnote{The term `dead
sequence' is not 
preferred because galaxies on the sloped `red sequence' are also dead in the
sense that they are not forming stars.} 
\citep{2008MNRAS.389...13R, 2011MNRAS.417..785J} 
at the high-mass end of the `red sequence' is
observed and shown to be defined by the (pure) E galaxies
\citep{2022A&A...666A.170Q}, with the dust-poor S0
galaxies defining the sloped segment of the `red sequence'.  These somewhat known but
often missed elements of, and clues in, the CMD are evident here due to
the refined galaxy morphologies. This 
engenders a better understanding of galaxy evolution than can be gleaned from
using just an LTG versus ETG distinction. 

Indeed, some motivation for investigating the CMD stems from recent developments in
galaxy morphology studies. In particular, from the realisation that the dust
morphology, and likely dust content\footnote{The dust-to-gas ratio is
understandably higher 
in merger-built S0 galaxies than in S galaxies \citep{2008ApJ...678..804E, 2014MNRAS.444L..90B}.},
of S0 galaxies, tracks their origin \citep{Graham-S0}.
This revelation into dust-poor and dust-rich S0 galaxies
is likely also tied to the previously unexplained dual population of
(low- and high-mass) S0 galaxies detected by \citet{1990ApJ...348...57V}.
Furthermore, it 
also incorporates the findings of \citet{1992AJ....104.1039S} and
\citet{2008AJ....136....1W}, which revealed
that ETGs with blue colours, relative to the `red sequence', display signs of
merger activity leading to star formation and thus a reduced mean age and more blue
colour.  Applying this knowledge to the CMD lends to additional growth
trajectories beyond LTGs in the `blue cloud' turning into red spirals
\citep{1951ApJ...113..413S, 1972ApJ...176....1G} or (low mass) discs transforming into ETGs
\citep{1981ApJ...243...32F, 1996Natur.379..613M}, or LTG collisions directly (under ideal
collisions) producing E galaxies \citep[e.g.,][]{1979A&A....76...75R,
  1981MNRAS.197..179G, 1983MNRAS.205.1009N}.
Moreover, the analysis herein
rewrites the description from \citet{2014MNRAS.440..889S} for populating the
`green valley'. Rather than S galaxy mergers rapidly quenching the star formation
to produce red galaxies, these merger remnants tend to have persistent star formation \citep{Graham-SFR}
and are found here to linger at the green end of the `blue cloud'. 

Section~\ref{Sec_col_bh} presents the second application of the `Triangal',
revealing its presence in the colour-(black hole mass) diagram (CM$_{\rm
  bh}$D).  Relations for dust-poor and dust-rich galaxies are established.
Section~\ref{Sec_Disc} provides an extended discussion, while
Section~\ref{Sec_Sum} summarises the main points.

\section{Greater Context Setting: Historical developments}
\label{Sec_Hist}

While this section can be skipped, it provides a fuller introduction to galaxy
morphology, which some readers may appreciate.  With what is now a relatively
comprehensive schema for galaxy speciation, it seemed apt to review several
relevant aspects as to how we got here. This section helps to place ideas and
developments into a broader context than typically garnered, thereby providing
a deeper understanding of the generations of work leading to studies of the
CMD.  It is mentioned when different galaxy types were recognised, as are
past efforts to connect them via speculated morphological
changes. References are provided to works by many pioneers upon whose shoulders we now
stand.  This section then segues to relevant notes regarding
the measurement of galaxy colours.

\subsection{Galaxy Morphology}
\label{subMorph}

In the late 1700s and the first half of the 1800s, the Herschel family did not
call out any spiral nebula, and their $\sim$2500-member catalogues of
nebulae and clusters of stars identified none as such \citep{1786RSPT...76..457H,
  1789RSPT...79..212H, 1802RSPT...92..477H}.
It was not until Lord William Parsons, the Right Honourable
Third Earl of Rosse, built what was then the world's largest (72-inch)
telescope\footnote{At odds with John Herschel,
William Parsons was of the view that the `nebular hypothesis' for forming
planets around stars
\citep{Swedenborg1734, 1755anth.book.....K, LaPlace1796, LaPlace1825} 
was not the whole story and that many nebulae were, in fact, comprised of 
countless stars. His telescope was, in part, built to test this.} at Birr
Castle, Ireland, that 
a spiral pattern was identified \citep{1850RSPT..140..499R}.\footnote{An 
additional 256 non-stellar objects were discovered \citep{Parsons1861,
  Rosse1878} and would later be subsumed into the
New General Catalogue \citep[][]{1888MmRAS..49....1D}.}
Soon after, and perhaps for the first time,
\citet{1852AJ......2...95A} introduced the term `elliptical'  to distinguish
elliptical-shaped nebulae from the newly-found spiral nebula. 
John Herschel's General Catalogue of Nebulae and Clusters of Stars 
\citep{1864RSPT..154....1H} subsequently called out five spiral nebulae
among its $\sim$5,000 entries, 
which additionally included planetary nebula and globular clusters of stars. 
The New General Catalogue \citep[NGC:][]{1888MmRAS..49....1D} was an expanded compilation that 
contained thousands of entries.  While it also identified (only the same five)  2- and 3-branched
spiral nebulae, including the `magnificent spiral' in M51 reported by
\citet{1850RSPT..140..499R}, at that time, no 
nebulae were identified explicitly as lentil or lenticular in shape. 
This was also true for the subsequent Index Catalogues 
\citep{1895MmRAS..51..185D, 1910MmRAS..59..105D}. 
However, with the application of photography to astronomy \citep{1882MNRAS..42..367D,
  1882Natur..25Q.489., 1883MNRAS..43..255C},  
much deeper images could be acquired 
\citep{1893spss.book.....R, 1895MNRAS..56...70R}, 
and thus, fainter nebulae were discovered, and greater detail was observed in
those already found. 

While \citet{1920MNRAS..80..746R, 1920MNRAS..81..129R} mentions
`spindle'-shaped objects, which were thought to be edge-on spiral nebulae,
\citet{1925MNRAS..85.1014R} introduced examples of lenticular nebulae that are
lens-like or lentil in shape.  These objects were identified explicitly as the suspected
bridging population in the E to S evolutionary sequence of
\citet{1919pcsd.book.....J}, largely based on the `nebular
hypothesis' debated by J.\ Herschel and W.\ Parsons. 
The theory proposed that elliptical-shaped nebulae 
evolved into spiral-shaped nebulae by spinning-out material aided
by passing encounters \citep{1852AJ......2...95A, 1972ApJ...178..623T}.
Hence, the origin of the early- and late-type galaxy nomenclature used today.

Extrapolating upon the numbers of new nebulae seen in photographed regions of the sky using
the 8-inch Bache Doublet telescope at the Harvard College Observatory,
\citet{1890AnHar..18..113P}, with the aid of Mrs Williamina P.~S.\ Fleming,
estimated an increase of 50-60 per cent over the 7,840 nebulae in the NGC.
\citet{1899MNRAS..60..128K} later revealed there was far more to the cosmos
than ever realised before when he estimated there were over 120,000 nebulae
that would be seen if the sky was mapped with one-hour exposures using the
Crossley 36-inch reflector \citep{1900ApJ....11..325K}.
\citet{1904LicOB...3...47P} subsequently estimated a figure of 500,000 to one
million, and 
over the decades, this number has increased from nearly a billion through
photographic work \citep[e.g.,][]{1956VA......2.1607L} to 100-200 billion
that are bright enough to be seen in the observable portion of the Universe by
the {\it Hubble Space Telescope} \citep{2013ApJS..209....6I}, to a suspected two
trillion \citep{2016ApJ...830...83C} with $M_*>10^6$ M$_\odot$ out to $z=8$.
Due to galaxy mergers, these numbers can decline over time as the Universe matures
according to the evolving galaxy merger rate \citep{2016ApJ...830...83C}.

The question all this is leading to is, how do all these galaxies and galaxy
types fit together?  What clues may galaxy morphology (and colour) tell us
about their origin?  Galaxy classification schemes are an important first step
in addressing this, and several have been put forward over the years
\citep[see][for a review]{2019MNRAS.487.4995G}.  However, the challenging next
step has been deciphering the correct evolutionary pathways between the
different galaxy morphologies.  A solution materialises in Fig.~\ref{Fig0},
taken from \citet{Graham-triangal}, and it is pertinent to ask if and how the
transitions seen there appear in the CMD. This requires the addition
of key morphological information into the CMD beyond simply LTG versus ETG, and this is
done here.  The CMD is investigated for signatures of how galactic
metamorphism may play out.  To date, usually only the faded S galaxy and S galaxy
merger origin are considered for S0 galaxies
\citep[e.g.,][]{2020MNRAS.498.2372D, 2022MNRAS.515..201C}.
Although, \citet{2018ApJ...862L..12S} explored the fragmentation of unstable
discs at high-$z$, 
bypassing the spiral galaxy phase and leading to clumps that build up the bulge.

As for some of the better-known morphological classification schemes, Fig.~\ref{Fig_schemas}
provides a summary of how the `Tuning Fork' --- displaying what is arguably
still the most widely shown sequence of galaxy morphologies --- has evolved
over the last century to become the `Triangal' \citep{Graham-triangal}, with
evolutionary pathways and a new/key differentiation of lenticular galaxy
type. This has 
revealed the initially unexpected, but with hindsight, obvious, bridging nature of 
LTGs between dust-poor and dust-rich S0 galaxies. 
Skimming the schemas in Fig.~\ref{Fig_schemas}, it may be tempting to conclude
that the ES galaxies \citep{1966ApJ...146...28L} are E4 to E7 galaxies, but
this is not the case.  The elongated E4 to E7 galaxies are invariably
misclassified S0 
galaxies \citep[as noted by, for example,][]{1970SvA....14..182G,
  1984A&A...140L..39M}, while the ES galaxies contain intermediate-scale discs
that do not dominate the light at large radii, as is the case in S0 galaxies.
The subtype ES,e galaxies are akin to elliptical galaxies and are likely built from dry
major mergers of S0 galaxies that did not fully erase the disc component.
Some of 
Liller's ES galaxies have been labelled `disc ellipticals'
\citep{1988A&A...195L...1N}, and they are simultaneously fast and slow rotators
depending on the sampled aperture size \citep{2011ApJ...736L..26A,
  2017ApJ...840...68G, 2017MNRAS.470.1321B}.  They are also identified as S0$^-$ sp/E5-E7 type
galaxies in the scheme of \citet[][their table~1]{2015ApJS..217...32B}.
Advancing this work, \citet{Graham-S0} suggests that the more compact ES,b
galaxies \citep[aka `red nuggets':][]{2005ApJ...626..680D, 2009ApJ...695..101D} 
might have formed from more ancient mergers involving gas-rich galaxies from the 
high-mass end of what are today's dust-poor S0 galaxies.  The ES,b galaxies may have 
bypassed the S galaxy phase of evolution.  They appear more
akin to high-density bulges than low-density elliptical galaxies, hence the
introduced subscripts `b' and `e' on Liller's ES notation.\footnote{Perhaps
ES,b galaxies would be better called `lentical', and `ellicular' reserved for
the ES,e galaxies.}

\subsection{Colouring in the CMD}

When \citet{1936rene.book.....H} presented what has come to be known as the
Tuning Fork diagram ---  a rotated form of the \citet{1928asco.book.....J}
Y-shaped diagram,
as noted by \citet[][their footnote~42]{1971JHA.....2..109H} and
\citet{1997AJ....113.2054V}
---
he recognised that a galaxy's colour correlated with its morphology, such that
ETGs are red and LTGs are blue. 
It was subsequently realised that dwarf ETGs are bluer than giant ETGs
\citep[e.g.,][]{1959PASP...71..106B, 1961ApJS....5..233D,
  1978ApJ...225..742S}, collectively forming the `red sequence' of ETGs seen
in colour-(absolute magnitude) or colour-(stellar mass) diagrams
\citep{1961ApJS....5..233D, 1972MmRAS..77....1D}. This sequence flattens, and
the colours stabilise at both high masses \citep[e.g.,][]{2004ApJ...613..898T,
  2011MNRAS.417..785J} and low masses \citep[e.g.,][their
  figure~3]{1968ApJ...151..105D}; see also \citet[][their
  figure~10]{2017ApJ...836..120R}.\footnote{At these low masses, $M_{\rm
  *,gal} \lesssim 10^7$ M$_\odot$, the scatter in colour appears to become
quite unwieldy, and the sequence disappears in some works \citep[e.g.,][their
  figure~1]{2002ApJ...573L...5C}.}
In contrast, the LTGs do not display a strong
trend in the colour-magnitude diagram \citep{1972MmRAS..77....1D}, resulting
in their so-called `blue cloud'.\footnote{Interestingly, some of the galaxy
magnitudes and morphologies obtained in Australia by G\'erard and Antoinette
de Vaucouleurs came from the 30-inch Reynolds' telescope, which had been
donated by Reynolds in 1924 to what would become Mount Stromlo Observatory
\citep{2011AntAs...5...36M}.}

Well after the eventual fall from grace (by the mid-1920s) of
the `nebular hypothesis', \citet{1951ApJ...113..413S} still regarded the
transitional galaxy population between S and E galaxies as S0 galaxies.
However, they suggested this may occur from gas removal from, and the
subsequent fading and reddening of, S galaxies.  They speculated that a
collision of two S galaxies would be such that the stellar components would
pass through each other relatively unscathed. At the same time, the much higher number
density of the gas particles would result in encounters that leave the gas behind
at the collision site.\footnote{This phenomenon was also invoked to explain
the Bullet Cluster \citep{2008ApJ...679.1173R}.}
However, 
this idea tends to fail in the field environment because the stars and gas merge into a
single galaxy, and it is not applicable in galaxy clusters where the encounter
speeds are favourably higher but it is instead the 
cluster's hot halo of gas that washes out a galaxy's
cold gas \citep{1972ApJ...176....1G, 1973MNRAS.165..231D}, as can an array
of alternate fast-acting processes \citep[e.g.,][]{2004AJ....128.2677T, 
2006A&A...458..101A, 2018MNRAS.476.2137R, 2019MNRAS.487.3740W}. 

While rapid dust-removal, from, say, 
`ram-pressure stripping' \citep{1972ApJ...176....1G}, might initially act to make a galaxy
bluer, the associated gas-removal and the ensuing passive fading from stellar
evolution will soon redden a spiral galaxy and might result in the loss of the spiral
pattern \citep[e.g.,][]{2001MNRAS.323..839S, 2002ApJ...577..651B, 2006AJ....132.2432L,
  2008ApJ...674..742B, 2022MNRAS.513..389R}. 
The S galaxies can also slowly fade due to the 
gradual depletion of gas associated with star formation 
\citep[e.g.,][]{2007A&A...470..173B, 2013ApJ...772..119L}. 
However, the notion that {\it all} S0 galaxies are faded 
S galaxies was ruled out by the existence of S0 galaxies with higher
masses than spiral galaxies \citep{2005ApJ...621..246B}.
That notion implicitly presupposes that spiral galaxies
are primordial rather than having formed from 
pre-existing disc-dominated S0 galaxies. 

The `Triangal' takes the view that primordial 
star-forming S0 galaxies have faded to become
the $z=0$ dust-poor S0 galaxies; that is, there is no assumption that they
were once S galaxies. 
Furthermore, while the passive evolution and fading of an S galaxy would
change its colour from blue to red and therefore be associated with an upward
movement in the CMD from the `blue cloud' to the `red sequence', the fading
scenario encounters opposition when examining the $M_{\rm bh}$-$M_{\rm
  *,gal}$ diagram.  This is because the fading scenario, in which the gas is
gone and $M_{\rm *,gal}$ does not change greatly\footnote{Mass loss from
stellar winds may account for up to half the stellar mass.}, would require
$M_{\rm bh}$ to grow by an order of magnitude if faded S galaxies become 
dust-poor S0 galaxies (see Fig.~\ref{Fig0}). 
A related challenge for the fading S galaxy scenario is to grow the $M_{\rm
  bh}/M_{\rm *,sph}$ ratio by a factor of a few to match the $M_{\rm
  bh}$-$M_{\rm *,sph}$ relations for S and dust-poor S0 galaxies.

It is well known that, due to the relatively quick demise of massive hot blue stars and the longevity of
less massive red stars, the mean colour of a stellar population reddens as it ages.
This reddening occurs even in the presence of an exponentially declining star
formation rate \citep{1959ApJ...129..243S}. 
However, other factors are at play besides colour-changing
stellar evolution as stars progress through the Hertzsprung-Russell
\citep[HR:][]{1911POPot..22A...1H, 1914PA.....22..275R}
diagram \citep[e.g.,][]{1997ApJ...476...28P}.  For example, higher metallicity alters the
initial mass function \citep[IMF:][]{1955ApJ...121..161S, 2023MNRAS.tmp.3452T} 
of stars and retards the escape of energy from 
stars.  The metals in a star increase the opacity and thus the
internal pressure (thereby reducing the required luminosity and effective
temperature to balance gravity), which makes them appear redder. In
contrast, metal-poor stars are more blue.

An additional factor that is not intrinsic to the colours of a stellar
population is the dust clouds that screen/filter the light before it
reaches us, scattering the blue light and making the conglomerate of stellar
blackbody spectra from a galaxy appear redder than it is. Corrections for
Galactic extinction (through sightlines out of our galaxy) are readily applied
using infrared-based dust extinction maps \citep{1998ApJ...500..525S,
  2011ApJ...737..103S}. Empirical corrections for dust internal to 
external galaxies have been made from knowledge of the inclination of their
disc to our line of sight \citep{2008ApJ...678L.101D}. This also includes a
correction for dust when the external discs are viewed face-on. 
Roughly half of a spiral galaxy's bulge light at blue wavelengths,
coming from the far side of a face-on galaxy, does not reach us.  
Because of the central concentration of dust, bulge light is more obscured
than disc light. Given the trend of increasing bulge-to-disc, $B/D$, ratio with
the galaxy mass, external dust corrections have the effect of making the slight
slope to the `blue cloud' less steep; that is, when applied, this
dust correction is such that galaxies with higher $B/D$ ratios become more
blue than those with smaller $B/D$ ratios.

Refined measurements have brought better recognition of a bridging population of
galaxies in the CMD's so-called `green valley'.\footnote{The term "Green
Valley" was an off-the-cuff reference to Arizona's Green Valley retirement
village, implying that the `green valley' in the CMD is where S galaxies go
to retire, before becoming `red and dead'.} 
Some of these `green valley' galaxies have been 
identified as disturbed S0 galaxies \citep{1992AJ....104.1039S}, and some
were also identified as star-forming
\citep[][their figure~1]{2017ApJ...836..120R}. 
Among other things, this paper explores how merger-built S0 galaxies may
contribute to the `green valley' population --- specifically, involving S0
galaxies {\it a priori} known to be built
through gas-rich major mergers involving S galaxy collisions
\citep{1983MNRAS.202...37S, 1988MNRAS.234..733B, 1988Natur.335..705B,
  1996ApJ...471..115B, 2003ApJ...597..893N, 2012ApJS..203...17R}, as opposed
to fading S galaxies.  Another important distinction that needs
emphasis here is that this concept of dust-rich S0 galaxy formation
differs from the idea of S
galaxy mergers building (pure) {\it elliptical} galaxies.

\section{Data}
\label{Sec_data}

\subsection{Galaxy sample and the presence of dust} 
\label{Sec_sample}

\subsubsection{Galaxies with $M_{\rm bh}$ measurements}
\label{Sec_sample_mbh}

Building on \citet{2016ApJS..222...10S}, \citet{2019ApJ...873...85D}, and
\citet{2019ApJ...876..155S}, the primary sample is presented in
\citet{Graham:Sahu:22a}, which tabulates 
the stellar masses of 104 galaxies and their directly measured black hole masses. 
This  sample covers a range of galaxy morphology, as discussed in \citet{Graham-triangal}.
In general, the E, S, and S0 galaxies were identified as such from a combination of
visual inspection and multicomponent decomposition with recourse to kinematic
profiles and maps when available. 
Going beyond S\'ersic-bulge $+$ exponential-disc fits
\citep[e.g.,][]{1995MNRAS.275..874A, 1998MNRAS.299..672S, 2000ApJ...533..162K,
  2001AJ....121..820G}, 
the core-S\'ersic model was employed when needed, and the disc model could 
be either truncated, anti-truncated, or inclined rather than always exponential.
Furthermore, building on the pioneering work of \citet{1996A&AS..118..557D}
and \citet{1997AJ....114.1413P}, 
additional components such as bars, which are often more massive
than the bulges \citep{2007MNRAS.381..401L}, were modelled as a separate component. 
So, too, were nuclei, ansae, and rings, which can be low in mass yet bias the
fitted bulge model due to their significant surface brightness over a
small region.
Furthermore, radial variations in the isophotal contours, as traced by Fourier
Harmonics \citep{1978MNRAS.182..797C, 1990ApJ...357...71O,
  2000A&A...361..841A, 2015ApJ...810..120C, 2016MNRAS.459.1276C} 
were used to help distinguish (peanut shell)-shaped structures and inner
discs, sometimes referred to as pseudobulges, that differ from bulges.

The ES
galaxies have intermediate-scale discs rather than the large-scale discs that
dominate at large radii in S0 galaxies. 
Collectively, the above resulted in improved bulge magnitudes that were 
used in conjunction with the refined morphologies to uncover the 
morphology-dependent black hole mass scaling relations shown in 
Fig.~\ref{Fig0} and explained in \citet{Graham-triangal}.
This level of sophistication reveals critical departures from the linear
$M_{\rm bh}$-$M_{\rm *,sph}$ relation suggested by
\citet{1988ApJ...324..701D}. 

The proximity of the sample means that
there is little, albeit some, scope for misclassification of the galaxy morphology. Of particular note
is that the S0 galaxies are not treated as a single population but separated
according to four `dust bins', encompassing:
\begin{itemize}
\item no dust, dust=(N)o, 
\item  not much, only a nuclear dust ring/disc, dust=(n)uclear, 
\item  some weak/wide(r)-spread dust, dust=(y)es,
and
\item strong dust presence, dust=(Y)es. 
\end{itemize} 
\citet{Graham-S0} reports the presence/absence of dust in the S0 galaxies.
The literature also reported that those S0 galaxies, which are dust-rich, formed from
a major wet merger, while those that are dust-poor did not.
Small bulges and undisturbed stellar morphologies are 
characteristics of these dust-poor S0 disc galaxies.

This sample, with known black hole masses, is used to revisit the
colour-(black hole mass) diagram CM$_{\rm bh}$D in Section~\ref{Sec_col_bh}.
For the CMD (Sections~\ref{Sec_CMD} and \ref{Sec_corr}), this sample is
complemented with LTGs and ETGs from the Virgo Cluster.

\subsubsection{Virgo Cluster LTGs}

A sample of 77 (mostly, see below) LTGs in the Virgo Cluster with SFRs
greater than $\sim$0.3 M$_\odot$ yr$^{-1}$ is used.  The sample is comprised of 
74 LTGs from \citet{2019MNRAS.484..814G} supplemented with 
NGC~4407, NGC~4492, and NGC~4496a from \citet{2022MNRAS.512.3284S}, which
details the full sample.
An array of optical images of the 77 Virgo Cluster LTGs were checked for dust in case
ram-pressure stripping had removed it, as is occurring in the (not in sample) S galaxy
NGC~4522 \citep{1999AJ....117..181K, 2000A&A...364..532V}.
This is perhaps unlikely given the SFRs.  Nonetheless, 
the internal dust corrections are not applicable if the dust has been removed. 
However, it turns out that the sample is dust-rich, and they
also have predominantly low-to-zero bulge-to-total stellar mass
ratios (established from 3.6~$\mu$m
images, analysed by the Spitzer Survey of Stellar Structure in Galaxies
(S$^4$G) \citep{2010PASP..122.1397S} project\footnote{
%
%
\url{https://irsa.ipac.caltech.edu/data/SPITZER/S4G/}
}).  
Some discs,
for example, in NGC~4607, appear sufficiently edge-on that it cannot be
discerned if a spiral pattern is present.  Nonetheless, they are all dust-rich
and require a correction to their optical luminosity. 

Among the sample, five galaxies stood out.
NGC~4429 is an S0
galaxy with a figure-of-eight pattern and (only) a nuclear dust disc, while
NGC~4469 is a more dusty counterpart to NGC~4429. 
NGC~4457 is a dusty S0/S transition
galaxy developing a stellar spiral arm, as may be the dust-rich peculiar S0
galaxy NGC~4606, while NGC~4492 represents a slightly earlier phase with dusty
spiral arms visible at optical wavelengths but not yet sufficiently developed
to show up in the S$^4$G 3.6~$\mu$m image.  The above five galaxies are classified
here as S0 rather than S galaxies.  Corrections for dust in these galaxies are
applied. However, stellar masses (Section~\ref{Sec_mass})
were unavailable for NGC~4429 and NGC~4469 
and for one additional S galaxy: NGC~4647.  This reduced the working sample to
74, which includes three S0 galaxies.  A further three LTGs (NGC: 4303; 4388;
and 4501) were removed
because they were already in the sample with directly measured black hole 
masses.  Thus, the sample size of additional LTGs is 68, with an additional three S0 galaxies.

\subsubsection{Virgo Cluster ETGs}

The Virgo Cluster has been a popular target for constructing CMDs
\cite[e.g.,][]{1969AJ.....74..354T, 1977ApJ...216..214V, 1978ApJ...223..707S,
  1992MNRAS.254..601B, 2008AJ....135..380L, 
  2013A&A...552A...8D, 2013ApJ...772...68S} 
Here, a sample of 100 ETGs in the Virgo Cluster with {\it Hubble Space
  Telescope} ACS imaging \citep{2004ApJS..153..223C} 
Each S0 galaxy needed to be assigned to one of the four previously
mentioned `dust bins' (Section~\ref{Sec_sample_mbh}). This task was made easy as 
the appearance of dust was already noted by \citet{2006ApJS..164..334F}. 
Deviating from \citet{2006ApJS..164..334F},
the possible faint dust filament in the low-mass galaxy 
IC~3468 ($M_* \approx 2\times10^9$ M$_\odot$) remains elusive and has not been
counted. Signs of dust in the rest of the ETG sample were searched 
for and possibly found in three low-mass galaxies: IC~3025, IC~3492, and VCC~1512, 
all of which are recognised dIrr/dE transition galaxies with $(g-i)_{\rm AB} < 0.8$ mag. 

A few of the 
massive ($M_* > 10^{11}$ M$_\odot$) ETGs have rather weak dust features: 
NGC~4472 (E/S0, weak lane);  
NGC~4486 (E, weak filaments); 
NGC~4406 (S0, weak filaments); and 
NGC~4382 (S0,pec., weak patches).  
Although weak filaments may be a sign of minor accretions
\citep{2012MNRAS.419.1051L}, 
the dust patches in NGC~4382 are regarded in the literature as a fading signature of its 
merger origin, evident by its stellar shells, ripples, and plume/tail 
\citep{1988ApJ...328...88S, 1988AJ.....95..422E, 1994cag..book.....S}. 
The shells around NGC~4472 \citep{2016ASSP...42..165S} are also indicative of
its merger origin, as is its depleted stellar core \citep[][and references
  therein]{2014MNRAS.444.2700D} 
and bimodal globular cluster system \citep{1996AJ....111.1529G}. 
Once these massive galaxies are immersed in a hot, X-ray-emitting, gas halo \citep{2003ApJ...599...38B}, 
the dust is effectively evaporated, returning the metals to the gas phase
\citep{1979ApJ...231...77D, 2008MNRAS.387...13K}.  

Two low-mass, dust-rich ETGs are also worthy of comment: 
VCC~1250 (NGC~4476) and VCC~571. 
The strong $\sim$2~kpc dust lane/disc in NGC~4476 ($M_* \approx 10^{10}$ M$_\odot$) 
is associated with $10^8$ M$_\odot$ of molecular hydrogen 
having an external origin \citep{2000AJ....120..123T, 2002AJ....124..788Y}. 
The galaxy is regarded as having experienced a merger event, presumably involving the acquisition 
of a less massive, gas-rich spiral galaxy, which has led to the different angular momentum of
the gas from the bulk of the stars.
In general, ETGs with prominent dust lanes tend to be known merger remnants 
\citep[e.g.,][]{2012MNRAS.423...49K, 2020ApJ...905..154Y}.
The other dwarf galaxy, VCC~571 ($M_* \approx 10^{9}$ M$_\odot$), displays a young blue core
\citep{2017A&A...606A.135U, 2019A&A...625A..94H} 
and is regarded as a transition-type dwarf galaxy by \citet{2013MNRAS.436.1057D}. 

Of the 100 ETGs, nine (NGC: 4374; 4434; 4473; 4486; 4552; 4578; 4621; 4649;
and 4762) are already included among the sample with directly measured black
hole masses.  They are, therefore, removed to give a sample of 91 additional
ETGs for use in the CMD.  Of these, three are considered here to be dust-rich
S0 galaxies (NGC~4476, NGC~4526, and VCC~571).

\subsection{Galaxy colours and stellar masses}\label{Sec_mass}

\subsubsection{Galaxies with $M_{\rm bh}$ measurements}

The sample with black hole masses had their stellar masses derived by \citet[][their
  equation~4]{Graham:Sahu:22a} using {\it Spitzer Space Telescope}
\citep[SST:][]{2004ApJS..154....1W} Infrared Array Camera
\citep[IRAC:][]{2004ApJS..154...10F} 3.6~$\mu$m
luminosities coupled with the following expression for colour-dependent
mass-to-light ratios:
\begin{equation}
\log(M_*/L_{3.6}) = 1.034(B-V)_{\rm Vega} - 1.067, 
\label{Eq_MonL}
\end{equation}
valid for $0.5 < (B-V)_{\rm Vega} < 1.1$ mag.
This equation effectively assumes a diet-Salpeter IMF, as used by
\citet{2001ApJ...550..212B}, and therefore --- 
see \citet[][p.306]{2003ApJS..149..289B} and \citet[][their table~2]{2010MNRAS.404.2087B} ---
it yields logarithmic masses that
are 0.15~dex greater than would be obtained assuming a
\citet{2002Sci...295...82K} IMF.\footnote{Note: equation~4 from
\citet{Graham:Sahu:22a} was mistakingly derived by adding, rather than
subtracting, (0.3$-$0.225=) 0.075. As such the stellar masses 
and black hole scaling relations presented there are calibrated to the diet-Salpeter IMF.}
The $(B-V)_{\rm Vega}$ colours came from `The Third Reference Catalog' 
\citet[RC3:][]{1991rc3..book.....D}, as made available via
the {\it NASA/IPAC Extragalactic Database}
({\it NED}).\footnote{\url{http://nedwww.ipac.caltech.edu}}
They were corrected for Galactic extinction using the dust maps from
\citet{2011ApJ...737..103S}, also taken from {\it NED}. 
Eq.~\ref{Eq_MonL} stems from the $(B-V)_{\rm Vega}$-dependent
expression for the $K$-band $M_*/L_K$ ratios given by 
\citet[][their table~6]{2013MNRAS.430.2715I}.
It was based on realistic dusty models, designed
for ``samples that include a range of morphologies, intrinsic colours and 
random inclinations''.  
For blue LTGs with $(B-V)_{\rm Vega}=0.7$ mag, Eq.~\ref{Eq_MonL} gives $M_*/L_{3.6} \approx 0.45$, while for 
red ETGs with $(B-V)_{\rm Vega}=0.9$ mag, one has $M_*/L_{3.6} \approx 0.73$. 
The stellar masses obtained from this expression match closely with those
obtained using expressions in 
\citet{2015MNRAS.452.3209R}, \citet{2019MNRAS.483.1496S}, and
\citet{2022AJ....163..154S}.

\subsubsection{Virgo Cluster LTGs}

\begin{figure}
\begin{center}
\includegraphics[trim=0.0cm 0cm 0.0cm 0cm, width=1.0\columnwidth,
  angle=0]{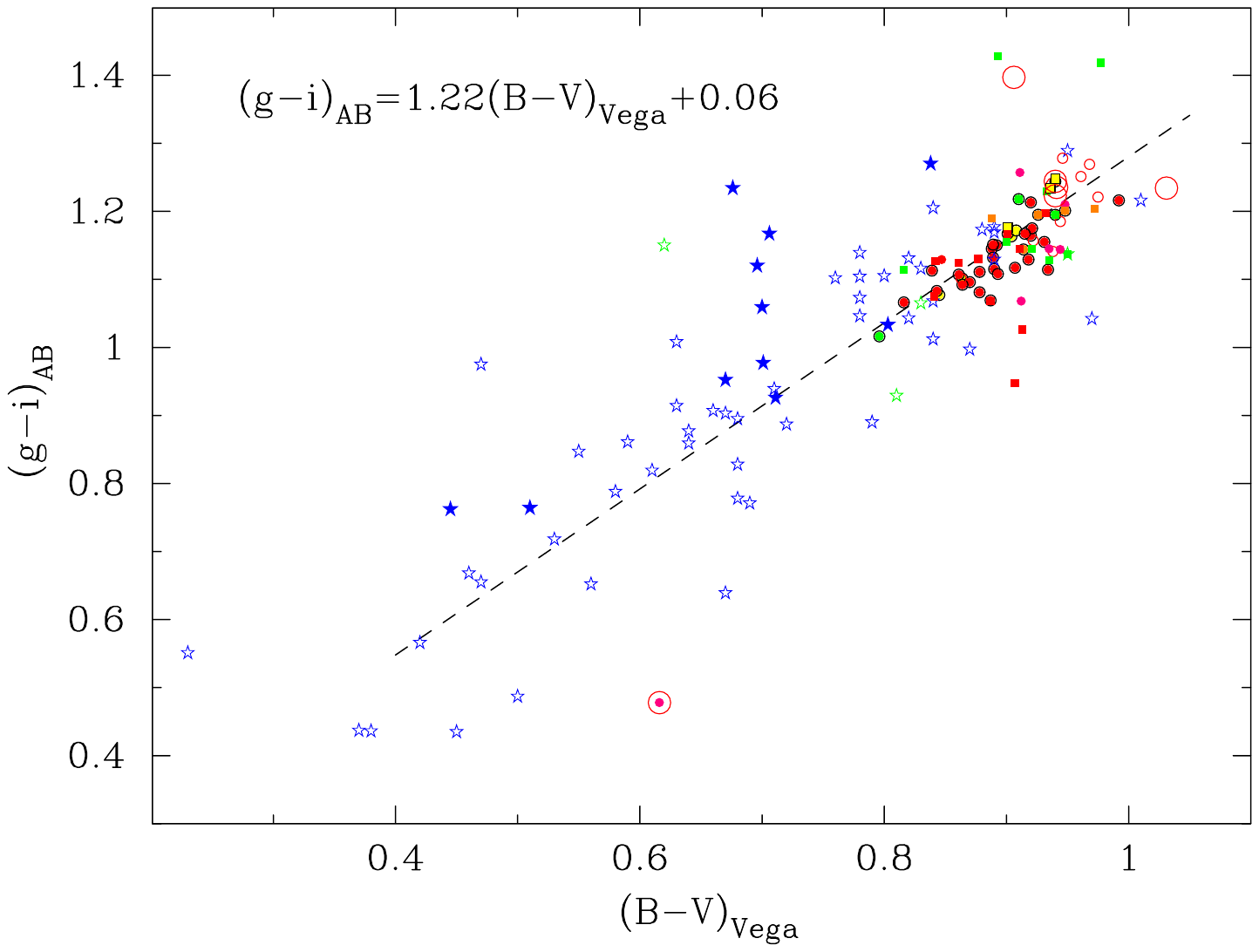}
\caption{(Galactic extinction)-corrected {\it SDSS} $(g-i)_{\rm AB}$ colour
  versus (Galactic extinction)-corrected {\it RC3} $(B_T-V_T)_{\rm Vega}$
  colour.  (The subscript $T$ denotes total galaxy magnitudes, but is dropped
  elsewhere for convenience.) ) 
  The dashed line is a simple approximation of the distribution. 
The ellicular (ES) BCG with $(g-i)_{\rm AB,obs}=0.48$ mag is NGC~1275, affected by an
AGN.  Symbols have the same meaning as in Fig.~\ref{Fig0}.  Also shown are
additional Virgo Cluster LTGs (open blue stars) plus three misclassified Virgo
Cluster S0 galaxies (open green stars, see Section~\ref{Sec_sample}), along
with additional Virgo Cluster ETGs denoted by the new circles that are
colour-coded according to the galaxy `dust bins': Y (green); y (yellow); n
(orange); and N (red). See Section~\ref{Sec_data} for details.  }
\label{Fig1}
\end{center}
\end{figure}

All but three of the Virgo Cluster LTGs had their stellar masses derived using 
S$^4$G 3.6~$\mu$m magnitudes\footnote{Available for all the Virgo Cluster LTGs, excluding NGC~4429,
  NGC~4469, and NGC~4647, the magnitudes were obtained at 
\url{http://cdsarc.unistra.fr/viz-bin/nph-Cat/html?J/PASP/122/1397/s4g.dat.gz}} 
coupled with distances from \citet[][their table~1,
  column~8]{2022MNRAS.512.3284S} 
and use of Eq.~\ref{Eq_MonL}. 
\citet{2023MNRAS.518.1352S} 
have shown that the 3.6~$\mu$m magnitudes measured by the S$^4$G team
were derived consistently with the 3.6~$\mu$m magnitudes obtained for the
sample with directly measured black hole masses \citep{2016ApJS..222...10S,
  2019ApJ...873...85D, 2019ApJ...876..155S}. However, 
$V$-band magnitudes were unavailable for
20 of the 68 LTGs in the Virgo Cluster, 
which meant that $(B-V)_{\rm Vega}$ colours were unavailable.
Of those 20 galaxies, 14 had
{\it Sloan Digital Sky Survey} \citep[{\it
    SDSS:}][]{2000AJ....120.1579Y}\footnote{\url{https://www.sdss3.org/}} 
$(g-i)_{\rm AB}$ colours that were corrected for Galactic extinction
\citep{2011ApJ...737..103S}\footnote{The $(g-i)_{\rm AB}$ colours and
Galactic extinctions were taken from {\it NED}.}
and used to estimate the $(B-V)_{\rm Vega}$ colour under the 
approximation seen in Figure~\ref{Fig1} and denoted by the relation 
\begin{equation}
(g-i)_{\rm AB} = 1.22(B-V)_{\rm Vega} +0.06. \label{Eq_col}
\end{equation}
To obtain the stellar masses for these 14 Virgo Cluster LTGs, 
the 3.6~$\mu$m luminosities were coupled with 
\begin{eqnarray}
\log(M_*/L_{3.6}) &=& 1.034[(g-i)_{\rm AB}-0.06]/1.22 - 1.067 \nonumber \\
&=& 0.848(g-i)_{\rm AB}-1.118. \label{EqThird} 
\end{eqnarray} 
For $(g-i)_{\rm AB} = 0.9$ mag, one obtains $M_*/L_{3.6}\approx 0.44$. 
Collectively, Eq.~\ref{Eq_MonL} and \ref{EqThird} 
provided stellar masses for an additional 62
(=68 minus 6 with no $(B-V)_{\rm Vega}$ nor $(g-i)_{\rm AB}$ colour) 
LTGs (plus 3 S0 galaxies) in the Virgo Cluster.

A caveat is that seven of these 62 galaxies have $(g-i)_{\rm AB}$ colours less than
$\sim$0.6 mag (see Fig.~\ref{Fig2a}),
indicative of a $(B-V)_{\rm Vega}$ colour less than $\sim$0.5 mag (see Fig.~\ref{Fig1}), 
thus giving $M_*/L_{3.6} \lesssim 0.3$. 
The $M_*/L_{3.6}$ ratio may not be reliable for these seven galaxies, such 
that the actual $M_*/L_{3.6}$ ratio may have been underestimated and more
likely be in the range of 0.3--0.4.
The models of \citet[][their figure~2]{2022AJ....163..154S} suggest that the
ratio may `bottom out' at around 0.36,
corresponding to $(g-i)_{\rm AB} = 0.8$ mag or $(B-V)_{\rm Vega} = 0.6$ mag.
\citet[][their figure~10]{2013MNRAS.430.2715I} reveals this effect with
the $M_*/L_{2.2}$ ratio. Given the bluer $(g-i)_{\rm AB}$
colours of these seven galaxies, this would imply that they may have had their
masses underestimated by up to a factor of two, which should be borne in mind
when looking at Figs.~\ref{Fig2a} and \ref{Fig2b}.

\subsubsection{Virgo Cluster ETGs}

The bulk of the Virgo Cluster ETGs do not have S$^4$G 3.6~$\mu$m magnitudes.
However, GOLD Mine\footnote{http://goldmine.mib.infn.it/} 
\citep{2003A&A...400..451G} 2.2~$\mu$m magnitudes are available for all but three 
of the 91 additional ETGs, and {\it SDSS} $ugriz$ colours are available for
all. This is fortuitous as 
$V$-band magnitudes were unavailable for this sample's fainter half.  
Therefore, the $(g-i)_{\rm AB}$ colour --- 
which tracks the $(B-V)_{\rm Vega}$ colour (Fig.~\ref{Fig1}, Eq.~\ref{Eq_col}) --- 
has been used.

\citet{Graham:Sahu:22a} note that the equation
\begin{equation}
\log(M_*/L_K) = 1.055 (B-V)_{\rm Vega} - 1.066 \,(+ 0.075)
\label{Eq_22}
\end{equation} 
from \citet[][their table~3]{2013MNRAS.430.2715I}\footnote{The modification
here is the adjustemnt by 0.075 dex to switch from a
\citet{1998ASPC..134..483K} IMF to a diet-Salpeter IMF
\citep{2003ApJS..149..289B}.} --- from a simpler galaxy model --- might be more
applicable for ETGs.  \citet[][see their figure~1]{Graham:Sahu:22a} also
reveal that it matches well with the $(B-V)_{\rm Vega}$-dependent expression
for $M_*/L_K$ ratios given by \citet[][their table~6]{2013MNRAS.430.2715I} 
for the dusty galaxy model that was used to derive Eq.~\ref{Eq_MonL}.  For
$(B-V)_{\rm Vega} = 0.7$ mag, Eq.~\ref{Eq_22} gives $M_*/L_{2.2}\approx 0.56$;
for $(B-V)_{\rm Vega} = 0.9$ mag, one obtains $M_*/L_{2.2}\approx 0.91$.
Eq.~\ref{Eq_col} can be used to modify Eq.~\ref{Eq_22}, such that  
\begin{eqnarray}
\log(M_*/L_{2.2}) &=& 0.865(g-i)_{\rm AB} - 1.043. 
\end{eqnarray} 
For $(g-i)_{\rm AB} = 0.91$ mag, one obtains $M_*/L_{2.2}\approx 0.55$; 
for $(g-i)_{\rm AB} = 1.16$ mag, one obtains $M_*/L_{2.2}\approx 0.91$.

Section~\ref{Sec_CMD} presents the galaxies' (Galactic extinction)-corrected
$(g-i)_{\rm AB}$ colours plotted against their stellar masses. 
  In Section~\ref{Sec_corr},  these colours 
are additionally corrected for dimming due to dust in the galaxies.
The above expressions for $M/L$ are purportedly applicable to dusty  models and
a range of galaxy morphologies and colours.  The revised colours are not used
to adjust the stellar masses, although this could be explored in future work.

\subsection{Correcting for dust}

Using galaxy samples with different disc inclinations, 
\citet{2007MNRAS.379.1022D} 
utilised the turnover luminosity, $L^*$, of each sample's (bulge and disc) luminosity function
to track the average extinction as a function of disc inclination.
Their work also included an adjustment for 
dimming due to dust in face-on discs.
These sample-averaged corrections are applied to individual galaxies
based on their disc inclination. 
Individual galaxies may, of course, have less or more dust
than the average disc galaxy.  Thus, these inclination-dependent dust corrections can be
too large or too small, respectively, when applied individually. 
However, when dealing with an ensemble of $N$ galaxies, as done here, this average dust 
correction works because the error in the mean shift to the cloud of points
declines with $\sqrt{N}$. 
\citet{2008ApJ...678L.101D} expanded the application to include the use of the
      {\it SDSS} $ugriz$ filter set, and the corrections given there are applied here to the 
S galaxies and the S0 galaxies with dust$=$Y \citep{Graham-S0}, including two
ES,b galaxies.\footnote{As noted in Section~\ref{subMorph}, ES,b galaxies 
seem more akin to S0 galaxies than E galaxies.}  

The dust corrections to a galaxy's bulge and disc magnitude, $\mathfrak{M}$,
can be combined to give the intrinsic galaxy magnitude from the observed
magnitude\footnote{In practice, $\mathfrak{M}_{\rm gal,obs}$ was first 
corrected for dust in our galaxy using the extinction maps of
\citet{2011ApJ...737..103S}.}, such that 
\begin{eqnarray}
\mathfrak{M}_{\rm gal,intrin} &=& \mathfrak{M}_{\rm gal,obs} -2.5\log \nonumber \\
&&  
\left\{ 
\left( \frac{B}{T} \right)_{\rm obs} 10^{\Delta \mathfrak{M}_{\rm bulge}/2.5}
+ 
\left[1-\left(\frac{B}{T}\right)_{\rm obs} \right]10^{\Delta \mathfrak{M}_{\rm disc}/2.5}
\right\}, 
\label{Eq_inc}
\end{eqnarray}
where 
$(B/T)_{\rm obs}$ is the observed
bulge-to-total luminosity ratio, and non-bulge components
such as bars and rings are effectively considered a part of the
disc.
The corrective terms are such that 
\begin{equation}
\Delta \mathfrak{M}_{\rm bulge} = \mathfrak{M}_{\rm bulge,obs} - \mathfrak{M}_{\rm bulge,intrin} =
b_1 + b_2 \left[ 1-\cos(i)\right] ^{b_3},
\label{Eq_Mb}
\end{equation}
and
\begin{equation}
\Delta \mathfrak{M}_{\rm disc} = \mathfrak{M}_{\rm disc,obs} - \mathfrak{M}_{\rm disc,intrin} = 
d_1 + d_2 \left[ 1-\cos(i)\right] ^{d_3},
\label{Eq_Md}
\end{equation}
where the coefficients $b_1$ to $b_3$ and $d_1$ to $d_3$ are
passband-dependent and given in \citet{2008ApJ...678L.101D}.  The (cosine of the) disc
inclination angle, $i$, is such that $i=90$ degrees for an edge-on disc and
$i=0$ degrees for a face-on disc.
This correction to the observed galaxy luminosity therefore requires the disc
inclination and the $B/T$ flux ratio.

For the galaxy sample with directly measured values of $M_{\rm bh}$, their
$(B/T)_{3.6}$ ratios are available from \citep{Graham:Sahu:22a}.
For that sample's LTGs, their disc inclinations are provided
by \citet{2017MNRAS.471.2187D}, and for the dust-rich S0 galaxies, 
$\cos\,i$ is roughly equal to the observed 
(projected on the plane-of-the-sky) minor-to-major axis ratio, $b/a$, at
the radii where the disc dominates.  More specifically, an average
ratio of vertical scale height, $z$, to radial scale length, $h$, i.e., disc thickness,
of 0.25 \citep[e.g.,][their figure~4]{1970ApJ...160..831S, 2014ApJ...787...24B} is assumed.
This ratio is also adopted for the three additional dust-rich S0 galaxies
in the Virgo cluster without a direct $M_{\rm bh}$ measurement, while 
a thickness of 0.21 \citep{2008MNRAS.388.1321P} is adopted 
for the Virgo Cluster's LTGs without $M_{\rm bh}$ measurements.
These $z/h$ ratios are used in the equation 
\begin{equation}
  \cos^2 (90-i) = \frac{1-(b/a)^2}{1-(z/h)^2}
  \label{Eq_thin}
\end{equation}
from \citet{1926ApJ....64..321H} to establish the inclinations of the discs
not listed by \citet{2017MNRAS.471.2187D}. 
For the Virgo Cluster LTGs without a directly measured black hole mass,
their observed $b/a$ axis ratios came from \citet{2010PASP..122.1397S}, and 
their 3.6~$\mu$m $B/T$ ratios were reported by 
\citet{2015ApJS..219....4S}.\footnote{\url{https://irsa.ipac.caltech.edu/data/SPITZER/S4G/overview.html},
see also \url{https://www.oulu.fi/astronomy/S4G_PIPELINE4/MAIN/}} 
For the Virgo Cluster ETGs without a directly measured black hole mass, there
are just three dust-rich galaxies requiring a dust 
correction. NGC~4526 ($b/a=0.33$), which resembles NGC~2787, has been assigned
$(B/T)_{2.2}=0.2$, while the lower-mass galaxies NGC~4476 ($b/a=0.68$) and
VCC~571 ($b/a=0.52$)
have been assigned $(B/T)_{2.2}=0.1$.

At 3.6~$\mu$m, Eq.~\ref{Eq_inc} is not required.  Indeed, rather than obscuring
the starlight, star-heated warm dust glows at 3.6~$\mu$m.  In LTGs, this 
typically accounts for $\sim$25 per cent of the 3.6~$\mu$m luminosity
\citep{2014ApJ...788..144M}. 
At 2.2~$\mu$m, the corrective term in Eq.~\ref{Eq_inc} is
typically $\sim$0.1 mag, and it is ignored here for the three dust-rich S0
galaxies from the Virgo Cluster ETG sample for which Gold MINE 2.2~$\mu$m
luminosities were used. 
Due to the higher concentration of dust in the centres of galaxies and the
blue colours of star-forming discs, the $B/T$ flux 
ratio decreases when we observe dusty disc galaxies at bluer wavelengths,
as seen in, for example, 
\citet[][figure~15]{2001AJ....121..820G},
\citet[][figures~5-6]{2004A&A...415...63M}, 
\citet[][figure~7]{2008MNRAS.388.1708G}, and more generally  
\citet{2014MNRAS.444.3603V}, \citet{2016MNRAS.460.3458K}, and 
\citet{2022A&A...664A..92H}. 
In the absence of multicomponent decompositions 
in the {\it SDSS} $ugiz$ bands for the current data samples,
the $(B/T)_{3.6}$ (and $(B/T)_{2.2}$ for three S0 galaxies)
ratios were simply reduced by a factor of 
1.0, 1.0, 0.5 and 0.25 to give the corresponding ratios in the 
$z$, $i$, $g$, and $u$ bands, respectively (see Section~\ref{Sec_Apdx} and
\citet[][figures~6]{2004A&A...415...63M} for the $u$-band). 
These ratios are used in Eq.~\ref{Eq_inc} to correct the $ugiz$ galaxy magnitudes for dust.

Among the sample with directly measured black hole masses, 
there are five dust-rich elliptical-like galaxies for which the dust 
corrections (designed for galaxies with large-scale discs) are problematic. 
These are comprised of one dust-rich elliptical galaxy (NGC~4374), two dust-rich ellicular
ES,e type galaxies (NGC~1275 and NGC~3607), and the two dust-rich ellicular ES,b
type galaxies (NGC~3115 and NGC~6861).
While NGC~4374 has a negligible SFR, NGC~3607 has been
measured to form stars at a rate of 0.25 M$_\odot$ yr$^{-1}$
\citep{Graham-SFR}, and arguably may require a dust correction.
NGC~1275 is very blue 
due to its AGN \citep{2001MNRAS.328..359I} and should be disregarded. 
In the past, ES,e and ES,b galaxies would tend
to be misclassified as (ordinary) E and (compact) S0 galaxies, respectively. 
Here, the dust correction is applied to just the two ES,b galaxies, but this
has little impact on the overall trends displayed by the larger sample.

\section{Analysis and Interpretation}
\label{Sec_Results}

\subsection{The observed CMD}
\label{Sec_CMD}

\begin{figure*}
\begin{center}
\includegraphics[trim=0.0cm 0cm 0.0cm 0cm, width=0.85\textwidth,
  angle=0]{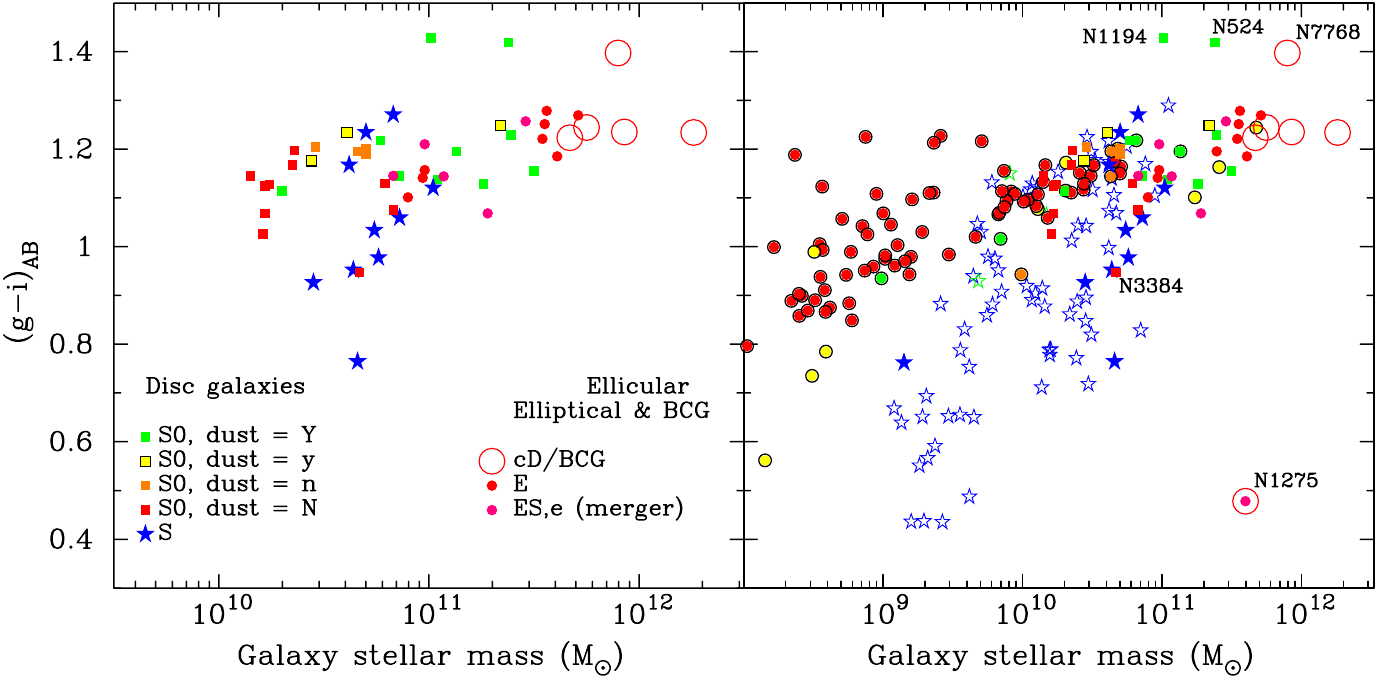}
\caption{Left panel: Colour-mass diagram for the sample having both directly
  measured black hole masses
  and {\it SDSS} colours (including the dusty galaxy NGC~1275 with an AGN). 
  Corrections for Galactic dust extinction have been made, but not for dust in
  the external galaxies. 
  Right panel: Addition of 62 Virgo
  Cluster LTGs (open blue stars) plus three misclassified Virgo Cluster S0 galaxies
  (open green stars, see Section~\ref{Sec_sample}) and 88 ($=100-9-3$)
  Virgo Cluster ETGs denoted by the new circles that are colour-coded 
  according to the galaxy `dust bins': Y (green); y (yellow); n (orange); and 
  N (red). See Section~\ref{Sec_data} for details. 
}
\label{Fig2a}
\end{center}
\end{figure*}

The $(g-i)_{\rm AB}$ colour versus stellar mass diagram (Fig.~\ref{Fig2a},
left-hand side)
resembles the $(B-V)_{\rm Vega}$ colour versus 3.6~$\mu$m
absolute magnitude diagram for the sample with directly measured black hole
masses \citep[][their
  figure~1]{Graham:Sahu:22a}.
This is indicative of the $(g-i)_{\rm AB}$
colour broadly tracing the $(B-V)_{\rm Vega}$ colour and the relatively narrow range of
$M/L_{3.6}$ ratios.  In the right-hand side of Fig.~\ref{Fig2a},
the CMD extends to lower-mass galaxies due to including the Virgo Cluster galaxies.
Reassuringly, the general pattern seen on the right-hand side of Fig.~\ref{Fig2a} for the ETGs
from the Virgo Cluster matches the $g^{\prime}-i^{\prime}$ CMD 
shown by \citet[][their figure~2]{2017ApJ...836..120R}.
The general trend also matches that seen elsewhere, such as reported by
\citet{2014A&ARv..22...74B} and \citet{2018AJ....155...69S}.
Thus far, only a correction for dust in our Galaxy
\citep{2011ApJ...737..103S}, courtesy of {\it NED}, 
has been applied to the optical colours in Fig.~\ref{Fig2a}. 
Therefore, one aspect which can be improved is to apply a correction for dust in the
external galaxies.

\subsection{The intrinsic CMD: Dust corrections in external galaxies}
\label{Sec_corr}

\begin{figure*}
\begin{center}
\includegraphics[trim=0.0cm 0cm 0.0cm 0cm, width=0.85\textwidth,
  angle=0]{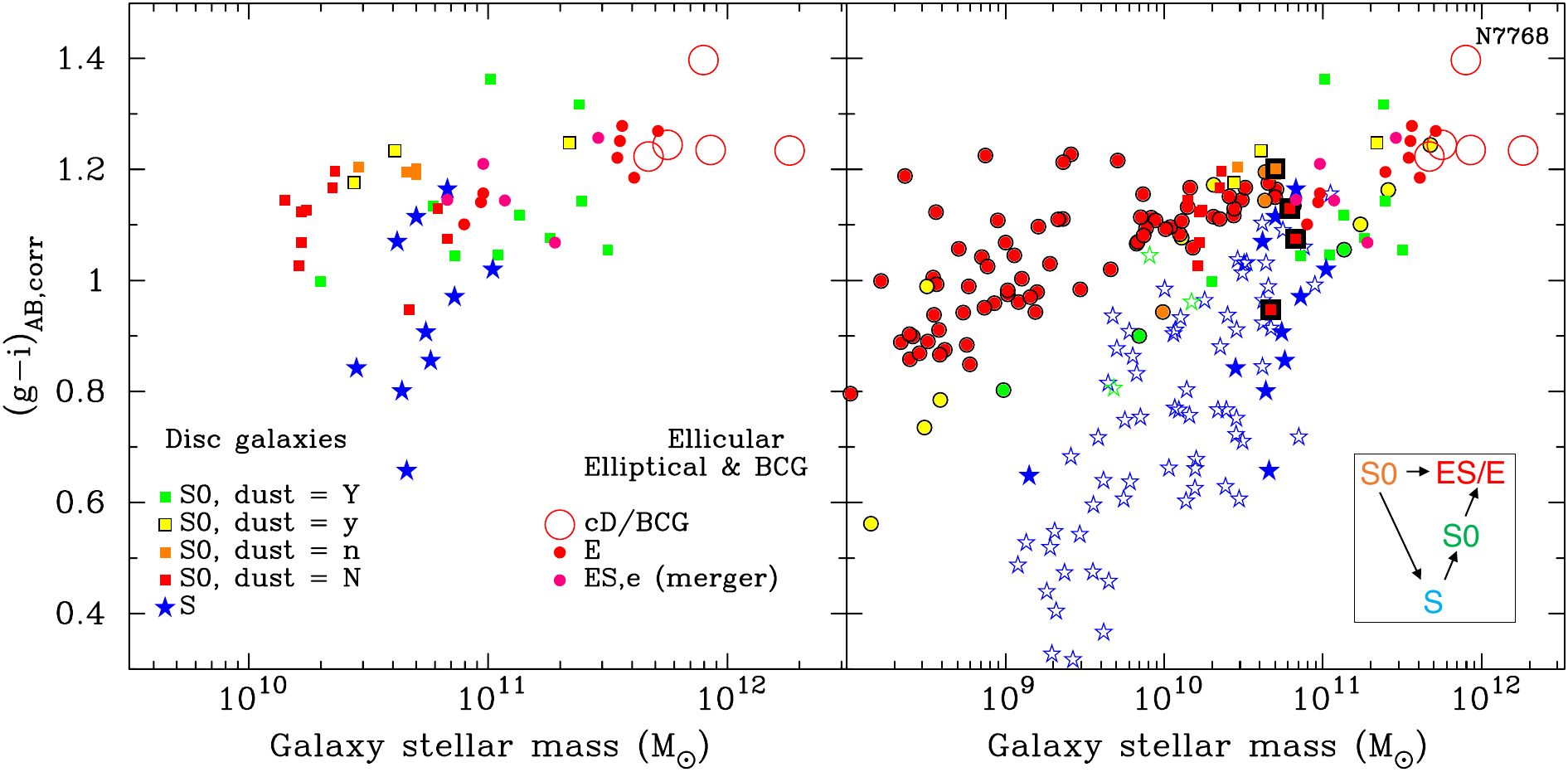}
\caption{Similar to Fig.~\ref{Fig2a}, except that the $(g-i)_{\rm AB}$ colour has
  additionally been corrected for dust in the external galaxies using the (disc
  inclination)-dependent prescription described in Section~\ref{Sec_corr} and following
  \citet{2008ApJ...678L.101D}.
  The AGN NGC~1275 has been removed. 
  Four dust-poor S0 galaxies that are possibly faded and transformed S
  galaxies are enclosed in a thick black square, contributing to the three
  types of S0 galaxy: primordial, wet-merger built, and faded S.  Spiral
  patterns emerge, often long ago, in what were initially spiral-less discs,
  and collisional weddings, i.e., major mergers, also drive galaxy
  speciation, captured by the `Triangal' \citep{Graham-triangal}.}
\label{Fig2b}
\end{center}
\end{figure*}

Fig.~\ref{Fig2b} reveals the shift in the CMD after correcting for dust. 
It illustrates that dust-corrected S0 galaxies, which contribute to the 'green
mountain', are co-located with S galaxies at the high-mass green end of the 'blue cloud' 
\citep{2022MNRAS.511.5405B}. 
These dust-rich S0 galaxies could be the population bolstering the `green
valley' seen by \citet[][their figure~11]{2007ApJS..173..342M,
  2007ApJS..173..293W} in the {\it GALEX} near-ultraviolet (NUV)-({\it SDSS})
optical CMD.

During wet (gas-rich) mergers of S galaxies, the gas particles, unlike the
stars --- which are separated by vast distances ---, have high number
densities, leading them to interact with each other. 
As a result, some gas particles lose speed and 
fall inward, while others get heated to high temperatures. 
The merging of these galaxies leads to the creation of central gas systems,
which may be associated with considerable AGN activity initially \citep{2000AJ....120..123T}. 
It is speculated here that this may also be the origin of some (tens to
hundreds of pc sized) nuclear discs in massive ETGs
\citep[e.g.,][]{2001AJ....121.2431R}; in contrast to the smaller, more
spheroidal, nuclear star clusters found in low-mass ETGs and LTGs
\citep[e.g.,][]{2002AJ....123.1389B, 2007ApJ...665.1084B, 2013ApJ...763...76S}.

\citet[][their
  figure~10]{2017MNRAS.471.2687B} suggest that the starbursting S0 galaxies
evolve rapidly from the `blue cloud' to the `green valley'.
\citet[][see their figure~~4]{2009ApJ...706L.173B} note that dusty starbursts can
appear in the `green valley' 
on the `red sequence' but that many shift back into the `blue cloud' after correcting for dust. 
\citet{2014MNRAS.440..889S}
also observed a probable S/S0 co-habitation/dual occupancy in the CMD,
where the bulk of their `green valley' sample of unknown 
``indeterminate'' galaxy 
type overlap with the high-mass green end of the S galaxies' `blue cloud'. 
However, while 
\citet{2014MNRAS.440..889S} speculated that these overlapping
unknown galaxy types were faded S galaxies,  
some/many may instead be merger products.
It is important to note that 
three of the four best candidates for faded S galaxies in the current sample reside on the `red
sequence' rather than still tapering off from the `blue cloud'. 
This favours rapid cold gas removal mechanisms acting on some S galaxies, 
at least in the present sample, to move them quickly through the `green valley' 
\citep[e.g.,][]{2009MNRAS.394.1991B, 2018MNRAS.477.4116K}, rather than 
gradual consumption of their cold gas via declining star formation 
effectively leaving them as S galaxies in the green end of the `blue cloud' \citet{2014MNRAS.440..889S}.

\begin{figure}
\begin{center}
\includegraphics[trim=0.0cm 0cm 0.0cm 0cm, width=1.0\columnwidth,
  angle=0]{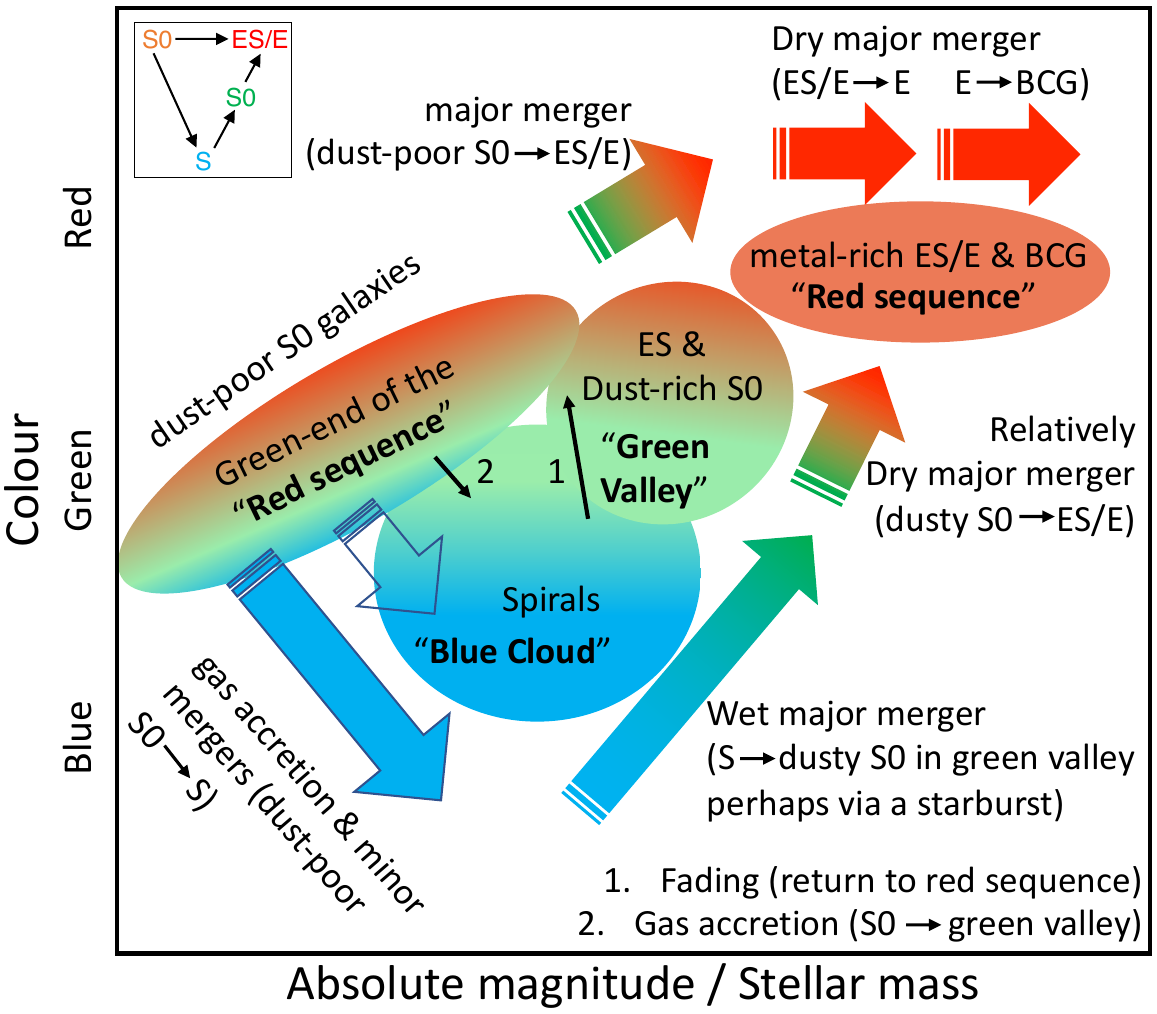}
\caption{Cartoon colour-mass diagram.  The large arrows show major
  evolutionary paths (punctuated equilibrium) taken by galaxies. The
  triangular-like structure introduced by \citet{Graham-triangal} and seen in
  Fig.~\ref{Fig_schemas} is evident.  Most S0-to-S transitions likely occurred
  in the past, when spiral patterns were induced in pre-existing discs, while
  the S-to-S0 merger-induced transitions are expected to occur in the past,
  present, and future. Most distant-future mergers are expected to be dry
  mergers.  Today ($z=0$), incomplete S0-to-S transitions may yield dwarf ETGs
  with blue cores. Tidal stripping may produce overly red dwarf galaxies for
  their (reduced) stellar mass, including compact E (cE) galaxies and tidal
  threshing may produce ultracompact dwarf (UCD) galaxies.  Some ES galaxies
  are closely associated with dust-rich S0 galaxies in the diagram; that is,
  there is some overlap rather than sharp dichotomies.  }
\label{Fig3}
\end{center}
\end{figure}

In Fig.~\ref{Fig3}, 
the large arrows denote the galaxy-merger-induced transformations. 
For those disc galaxies that did not develop a spiral pattern,
perhaps due to an early entry into a proto-cluster environment with a hot gas
halo, they would have become today's ($z=0$) dust-poor S0 galaxies. 
If gas acquisitions, gas recycling \citep[e.g.,][]{1991ApJ...376..380C}, and minor mergers are 
sufficient to keep a disc fuelled and cause gravitational
perturbations that produce/maintain a spiral pattern, then an S galaxy
arises \citep{1964ApJ...140..646L, 1966ApJ...146..810J,
  2013ApJ...766...34D}.  
However, the S0 $\rightarrow$ S passage is likely the domain of a bygone era when the Universe
was fuelling galaxies at a greater rate. 
Today ($z=0$), gas accretion onto a dust-poor S0 galaxy might only be
sufficient to 
move it into the `green valley' due to limited star formation 
\citep[e.g.,][]{2008MNRAS.385.1965H, 2009AJ....138..579K,
  2015Galax...3..192M}, in part because of
inclined, off-plane accretion \citep{2019ApJS..244....6S} and the high angular
momentum of the accreted gas \citep{2020MNRAS.491L..51P}.  This subdued level
of star formation 
is known as `rejuvenation' \citep{1990A&A...228L...9G, 1992AJ....104.1039S,
  1998A&A...333..419T, 2007MNRAS.381..245R, 2010ApJ...714L.171T,
  2015Galax...3..192M, 2022MNRAS.513..389R}. At $z=0$, this is expected to be more common in
the lower mass galaxies where the accreted material has a greater chance of
being a higher fraction of the galaxy's mass.  
This can be thought of as a kind of `downsizing'.  A reverse
(S $\rightarrow$ dust-poor S0) process involves the rapid (stripping) or slow
(consumption and exhaustion) removal of fuel, which can snuff out star formation and return a
blue galaxy to the `red sequence.'

Major merging of gas-rich S galaxies can result in a burst of star formation
and transform a pair of S galaxies into a dust-rich S0 galaxy along the `green
range'.  Rather than a `valley' or deficit between the `blue cloud' and `red
sequence', there (i) may be
something of a `mountain' or excess at $M_* \sim 10^{11}$ M$_\odot$ and (ii)
an additional population ranging to lower masses. Although more data would be
desirable, the dust-rich S0 galaxies appear to form a mid-to-low mass ridge on the green side of the
`red sequence' (Fig.~\ref{Fig2a}) and `blue cloud' (Fig.~\ref{Fig2b}).
The higher  abundance of merger-built dust-rich S0 galaxies at higher masses may
reflect a greater propensity for wet mergers to occur due to the stronger
gravitational attraction of more massive systems that are better able to turn
(otherwise flyby) encounters at a given velocity into mergers.

\subsection{The colour-(black hole mass) diagram: CM$_{\rm bh}$D}
\label{Sec_col_bh}

\begin{figure}
\begin{center}
\includegraphics[trim=0.0cm 0cm 0.0cm 0cm, width=1.0\columnwidth,
  angle=0]{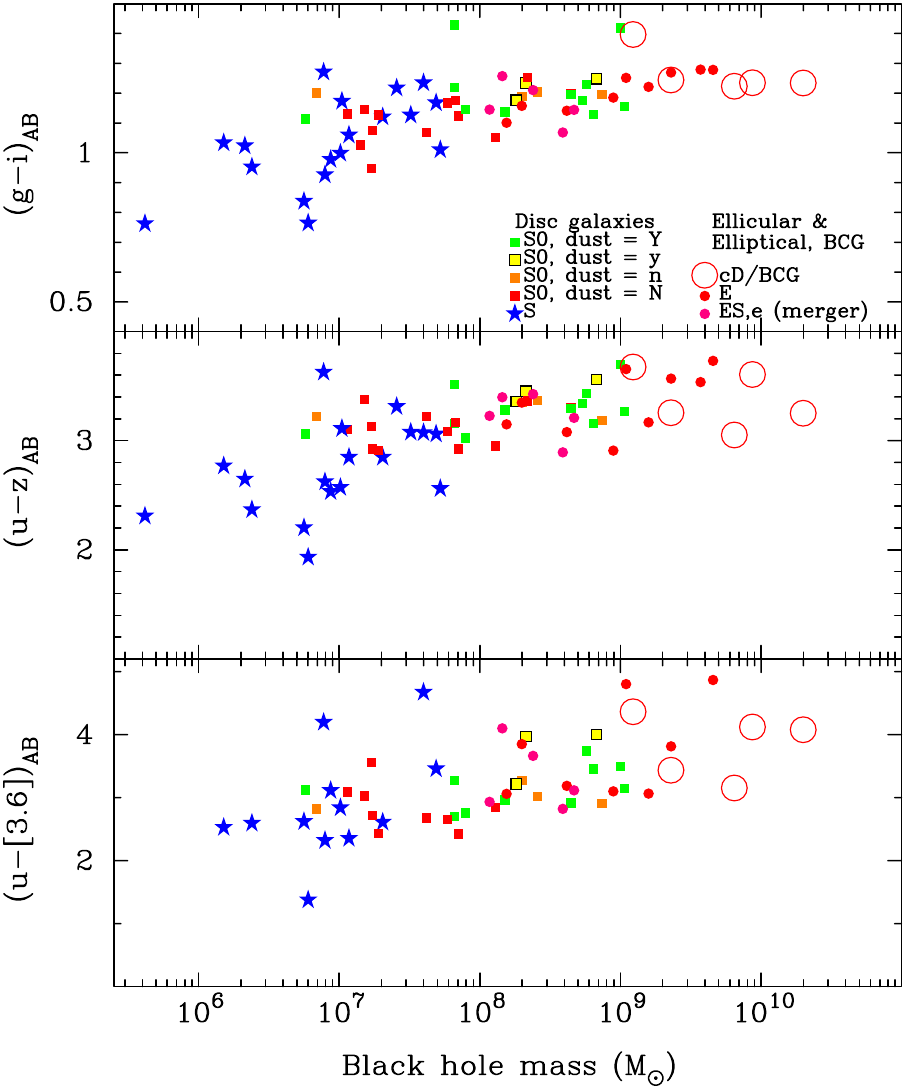} 
\caption{Colour-(black hole mass) diagram (CM$_{\rm bh}$D) for galaxies with
  directly measured black hole masses and available {\it SDSS} colours. 
Top panel: The reported separation between ETGs and LTGs seen in the $M_{\rm
  bh}$-$M_{\rm *,galaxy}$ 
diagram by \citet{2020ApJ...898...83D} is not (yet) evident here. The
distribution appears to asymptote in the red.
The three galaxies with $(g-i)_{\rm AB} \approx 1.4$ mag may be measurement errors.
While the Galactic extinction correction is applied here, no correction for
dust internal to the galaxies has been applied.
}
\label{Fig5}
\end{center}
\end{figure}

Given that galaxy masses are not required for the CM$_{\rm bh}$D, expanding the sample of
galaxies with directly measured black hole masses is possible.  Of the 145
galaxies with directly measured black hole masses listed in \citet{2019ApJ...887...10S},
70 have both $g$ and $i$ {\it SDSS}  magnitudes in the {\it NED} database,
and 69 have both {\it SDSS} $u$ and $z$ magnitudes.  There are 
59 having both {\it SDSS} $u$ and {\it Spitzer} [3.6] magnitudes.
These latter magnitudes encompass three galaxies with S$^4$G-derived [3.6] galaxy
magnitudes, with the remainder coming from \citet{2016ApJS..222...10S},
\citet{2019ApJ...873...85D}, 
\citet{2019ApJ...876..155S}, and
\citet{Graham:Sahu:22b}.
Fig.~\ref{Fig5} shows $M_{\rm bh}$ against the {\it SDSS} $(g-i)_{\rm AB}$ and $(u-z)_{\rm AB}$
colours and the {\it SDSS}-{\it Spitzer} ($u$-[3.6])$_{\rm AB}$ colour.
These colours are yet to be corrected for dust in the external galaxies; they
have only been corrected for dust in our galaxy. 

After \citet{2016ApJ...817...21S} suggested there was a red and blue sequence
in the $M_{\rm bh}$-$M_{\rm *,sph}$ diagram --- which has been supplanted by
the (galaxy morphology)-dependent relations seen in Fig.~\ref{Fig0} ---
\citet{2020ApJ...898...83D} introduced red and blue $M_{\rm bh}$-colour relations. That
study coupled {\it Galaxy Evolution Explorer} \citep[{\it
    GALEX:}][]{2007ApJS..173..682M} ultraviolet and {\it Spitzer} infrared magnitudes 
for 67 galaxies to provide FUV-[3.6] and NUV-[3.6] colours.
\citet{2020ApJ...898...83D} 
reported separate blue and `red sequences' for LTGs and ETGs (E, S0, and
S0/a).  This division is
not (yet) apparent in the current sample 
(Fig.~\ref{Fig5}). This is, in part,  likely due to the low number of S galaxies, in
particular at $10^7 < M_{\rm bh}/M_\odot < 10^8$. It is also, as shown next,
partly because of the dust in the external galaxies. 

\begin{figure}
\begin{center}
\includegraphics[trim=0.0cm 0cm 0.0cm 0cm, width=1.0\columnwidth,
  angle=0]{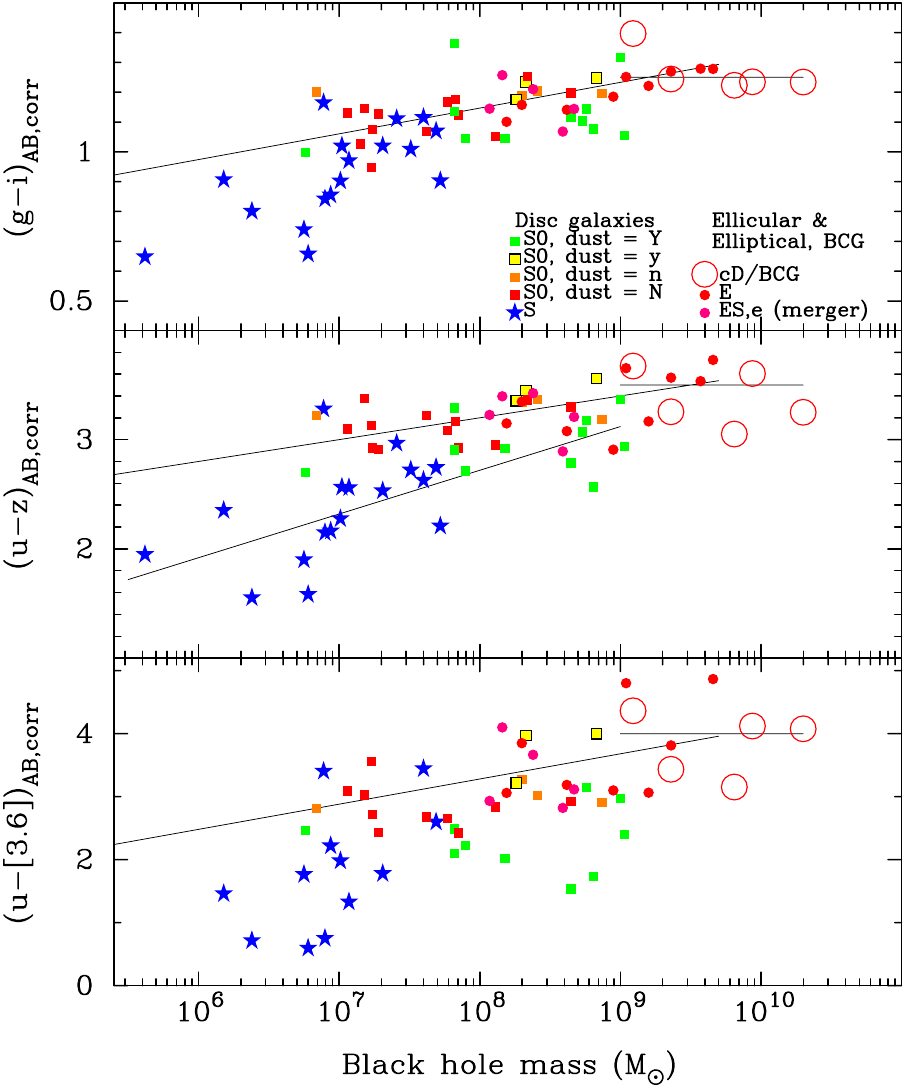} 
\caption{Similar to Fig.~\ref{Fig5} but with (disc inclination)-dependent
  dust-corrections applied to the S and dust-rich (dust$=$Y) S0 galaxies. 
The two red spirals with $(u-[3.6])_{\rm AB,corr} > 3$ mag are NGC~3368 and NGC~4258.
Parallels with the CMD (Fig.~\ref{Fig2b} and \ref{Fig3}) are evident, helping
to explain the CM$_{\rm bh}$D shown here, where one can detect a `red sequence' 
with a `red plateau' and a blue-green sequence/cloud with a green mountain/range.
Relations for the S and dust-rich S0 galaxies and for the dust-poor S0 and
ES/E/BCG galaxies are given by Eq.~\ref{Eqblue} and \ref{Eqred}--\ref{Eqreddest}, respectively.
}
\label{Fig6}
\end{center}
\end{figure}

\citet{2020ApJ...898...83D} used the dust
corrections from \citet{2008ApJ...678L.101D} to present the $M_{\rm bh}$-colour diagram, using
{\it GALEX} and {\it Spitzer} magnitudes for the colour.  Given that
\citet{2008ApJ...678L.101D} only provided
corrections from $u$ to $K$ [2.2~$\mu$m], \citet{2020ApJ...898...83D} adapted
the correction in the $u$-band for their {\it GALEX} UV data, and the
correction in the $K$-band for their {\it Spitzer} 3.6~$\mu$m data.
The approach taken here
differs slightly in a few other ways. 
First, while \citet{2020ApJ...898...83D} 
used the same corrective formula as \citet{2008ApJ...678L.101D}, they 
only applied it to the LTGs and not to the dust-rich S0 galaxies. 
This makes the LTGs' colours more blue relative to {\it all} the ETGs, 
enhancing the separation in colour.  A second (minor) difference is that \citet{2020ApJ...898...83D} 
used $\cos\, i = b/a$ to establish the inclination of the discs; that is, they
assumed the discs were infinitely thin (cf., Eq.~\ref{Eq_thin}).  
Third, they applied the dust-dimming correction to brighten 
the 3.6~$\mu$m magnitudes of the LTGs.  
For a disc-dominated LTG with a projected (on the plane of the sky) axis ratio
$b/a=0.25$, 
this amounts to a brightening of $\sim$0.18 mag. However, dust does not cause
dimming at 3.6~$\mu$m but instead glows.  

Fig.~\ref{Fig6} builds on the blue (star-forming) sequence and red (quiesced)
sequence shown in \citet{2020ApJ...898...83D}.  
Although it would be desirable to have a greater number of S 
galaxies at $10^7 < M_{\rm bh}/M_\odot < 10^8$ and more dust-poor S0 galaxies at
$M_{\rm bh} < 10^7 M_\odot$, as seen in Fig.~\ref{Fig2b}, 
the ETG/LTG separation seen by
\citet{2020ApJ...898...83D} when using the UV-[3.6] colour is evident in Fig.~\ref{Fig6}.
As suggested for the $M_{\rm bh}$-$M_{\rm *,gal}$
diagram (Fig.~\ref{Fig0}, right-hand side),
the S and dust-rich S0 galaxies appear to form a sequence when partnered. 
The blue sequence in \citet{2020ApJ...898...83D} is modified here to give a blue-green sequence that now 
includes S {\it and} dust-rich S0 galaxies, invariably recognised in the literature as
built from wet major mergers likely involving one or two S galaxies
\citep[][see the references therein]{Graham-S0}. Many of these dust-rich S0 galaxies
still have residual star formation \citep{Graham-SFR}.  The red`sequence is 
also modified here by excluding the dust-rich S0 galaxies and
recognising the near-constant colour for the (pure) E galaxies, creating a 
plateau at the high-mass end.

The $M_{\rm bh}$-$(g-i)$ relation shown in the
upper panel of Fig.~\ref{Fig6} for the 
`red sequence' (now excluding the dust-rich S0 galaxies) is given by 
\begin{equation}
  \log M_{\rm bh}=11.6[(g-i)_{\rm AB,corr}-1.0]+6.3, 
\label{Eqred}
\end{equation}
with a red plateau at $(g-i)_{\rm AB,corr}=1.25$ mag for $M_{\rm bh} \gtrsim
10^9$ M$_\odot$.
The corresponding $M_{\rm bh}$-$(u-z)$ relation shown in the middle panel is such that
\begin{equation}
  \log M_{\rm bh}=5[(u-z)_{\rm AB,corr}-3.0]+7.0, 
\label{Eqredder}
\end{equation}
with a red plateau at $(u-z)_{\rm AB,corr}=3.5$ mag.
Finally, the `red sequence' in the lower panel is given by
\begin{equation}
  \log M_{\rm bh}=2.5[(u-[3.6])_{\rm AB,corr}-3.0]+7.3, 
\label{Eqreddest}
\end{equation}
with a red plateau at $(u-[3.6])_{\rm AB,corr}=4.0$ mag.

The tentative relation for the blue-green sequence of LTGs and dust-rich S0
galaxies is less secure due to the dependence 
of these galaxy colours on the dust-inclination correction.
The following expression is shown in the middle panel of Fig.~\ref{Fig6}: 
\begin{equation}
  \log M_{\rm bh}=2.5[(u-z)_{\rm AB,corr}-3]+8.7. 
\label{Eqblue}
\end{equation}
Use of this blue-green relation to predict $M_{\rm bh}$ requires correcting the $u$ and $z$
magnitudes of one's disc galaxy (sample) for dust and disc inclination.
The $M_{\rm bh}$ versus FUV- and NUV-[3.6] colour relations for LTGs
in \citet{2020ApJ...898...83D} also require correcting for dust and disc
inclination, and knowledge of the $B/D$ flux ratio 
after excluding the flux of bars, which can be substantial.

Admittedly, 
the colour does not appear exceptionally useful for predicting $M_{\rm bh}$,
with a $\pm2\sigma$ scatter of around $\pm$1 dex for the red and blue
sequences.  However, this is competitive with recent $M_{\rm bh}$-$\sigma$
relations \citep{2016ApJ...818...47S, 2019ApJ...887...10S}. 
This is, however, an evolving area of research, and future refinements to
colours, from improved dust corrections, could alter the landscape.  The main
thrust of this subsection was to understand better the origin of the $M_{\rm
  bh}$-colour relations, rather than precisely define the relations, and to see
if and how the `Triangal' plays out in the CM$_{\rm bh}$D.

\section{Discussion}
\label{Sec_Disc}

This section discusses the rationale for and understanding of the
evolutionary pathways in the CMD. 
Simulations predict disc galaxies condensed from the gravitational collapse of
high angular momentum gas clouds \citep{1985Natur.317..595F,
  1993MNRAS.264..201K, 2007MNRAS.382..641S, 2008MNRAS.387..364Z}, 
and many redshift-zero low-mass ETGs are disc-dominated galaxies
\citep[e.g.,][]{1980MNRAS.193..189F}. 
They need not have ever hosted a spiral pattern.  As such, a high specific angular momentum
in galaxies today need not be evidence of a faded spiral galaxy
\citep{2018MNRAS.476.2137R}.  Models for spiral formation commence with a
perturbation in a pre-existing disc, and spiral patterns undoubtedly 
flourished in many of the initially  spiral-less discs, with ceers-2112 an example distant spiral at
$z\approx3$ \citep{2023Natur.623..499C}; see also A1689B11 at $z=2.54$
\citep{2017ApJ...850...61Y} and possibly BRI 1335-0417 at $z=4.41$
\citep{2021Sci...372.1201T}.

The $z=0$  stellar
masses of dust-poor S0 galaxies reach up to a few 10$^{10}$ M$_\odot$.  Substantial mergers and
accretions appear to have been inevitable above this threshold, resulting in an evolution of their
galaxy type (Fig.~\ref{Fig0}).
These low-mass ETGs can have younger (luminosity-weighted) stellar
populations than high-mass ETGs \citep[e.g.,][their
  figure~21]{2003AJ....125.2891C}. 
As such, they are not blue/green relative to high-mass ETGs solely because of
a lower metallicity \citep{2003AJ....125...66C}.  However, it must be noted
that most ($>$90 per cent by mass) of the stellar population in low-mass ETGs
is thought to be old \citep{2006AJ....132.2432L}.  Many are essentially a
primordial galaxy population, i.e., a first incarnation/generation, and they
can possess a 
low bulge-to-total ratio \citep[][figure~A2]{Graham-triangal}.\footnote{S0 galaxies with low
bulge-to-total stellar mass ratios are absent from the schematic in
\citet[][their figure~7]{2016ApJ...831..132G} because it 
implicitly focussed on `ordinary'/non-dwarf
galaxies with $M_{*,gal} \gtrsim 10^{10}$ M$_\odot$. This encompasses the
dust-rich S0 galaxies that typically have higher $B/T$ stellar mass ratios
than S galaxies \citep{2008MNRAS.388.1708G}.} 
This is also the case with the bulk ($>$75 per cent) of the stellar mass in
the bulges of S galaxies \citep{2009MNRAS.395...28M},
with star formation predominantly occurring in the disc where gas clouds 
cool and condense.  
By and large, these low-mass dust-poor S0 galaxies in the nearby Universe are metal-poor and may 
allow us to learn about the early Universe and test our theories of early disc
galaxy formation and growth \citep[e.g.,][]{1994MNRAS.267..401N,
  2008MNRAS.387...13K, 2011ApJ...740L..24K, 2011ApJ...731...54P, 
  2013MNRAS.436.2301P}.  Their mere existence today suggests that the galaxies at cosmic
dawn could have been spiral-less disc galaxies, and their $M_{\rm bh}/M_{\rm
  *,sph}$ ratios imply they are not faded S galaxies.

Disc galaxies with a sustained
fuel supply may have grown into S galaxies.  Minor mergers can build
bulges by bringing in material and transplanting disc stars, halo gas can cool,
and these mechanisms, along with star-forming turbulence, can produce gravitational perturbations that
invoke spiral formation \citep{1966ApJ...146..810J, 1996ssgd.book.....B}. 
The subsequent major merger of S galaxies can
destroy the spirals, transfer some disc and bar stars to the bulge component through
violent relaxation, but not fully erase the angular momentum \citep{1996ApJ...471..115B}, creating the
dust-rich S0 galaxies that tend to have more massive bulges and black holes
than the S galaxies (Fig.~\ref{Fig0}, left-hand side). The existence of these more
massive bulges in `green valley' galaxies was noted by
\citet{2018MNRAS.476...12B}.
Once they no longer look like mergers-in-progress, and the recent/current star
formation has sufficiently declined for the colour to change from blue to
green, detecting signs of mergers and interactions in galaxies beyond the local
($z\lesssim 0.03$) Universe becomes quite challenging. This may hamper some
studies' ability to detect the role of mergers in populating the `green valley'.

\begin{figure}
\begin{center}
\includegraphics[trim=0.0cm 0cm 0.0cm 0cm, width=1.0\columnwidth,
  angle=0]{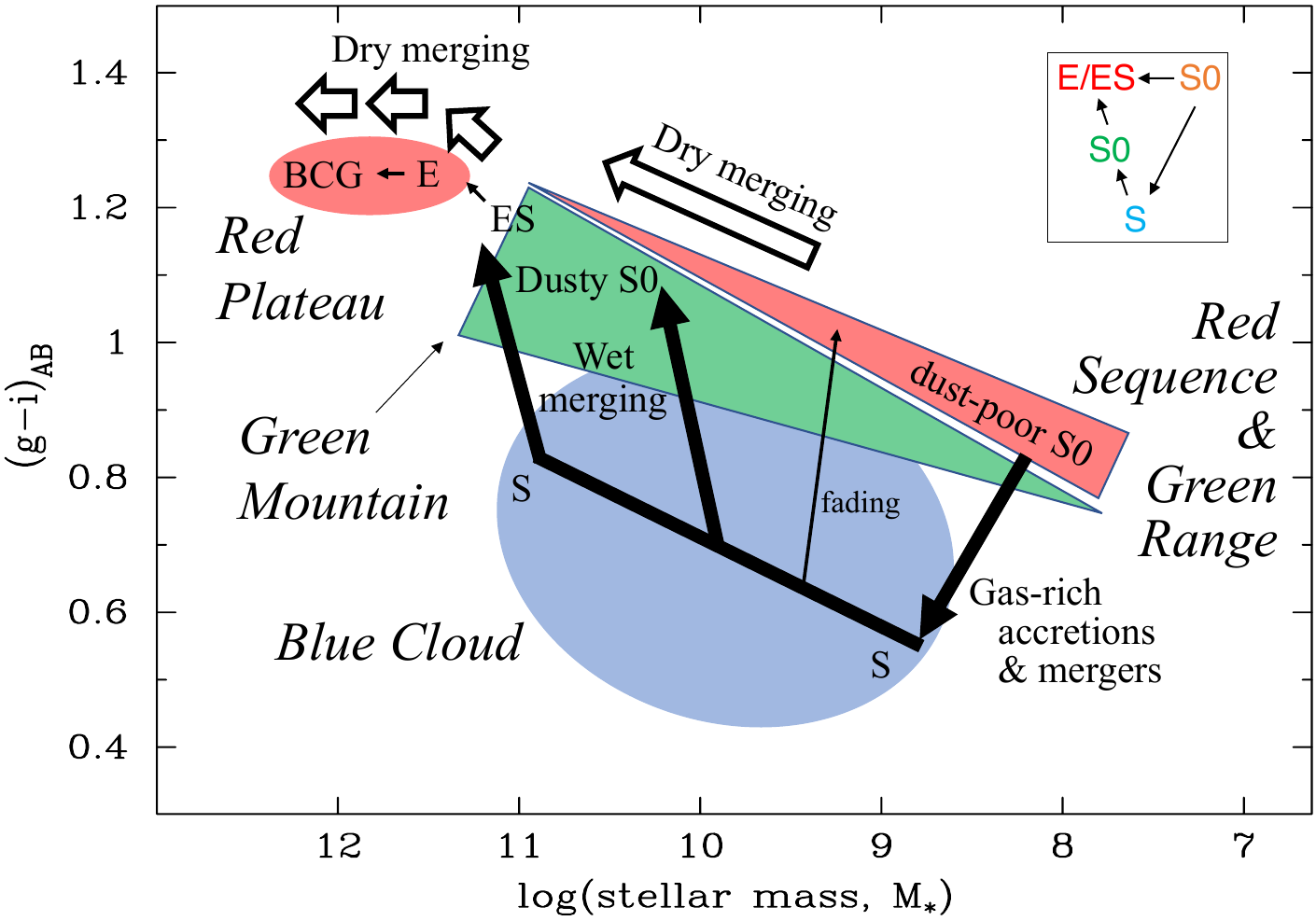}
\caption{Adaption of the colour-mass schematic shown in \citet[][their
    figure~10]{2007ApJ...665..265F}.  Today's $z\sim0$ dust-poor S0 galaxies
  were once blue, but they need not have ever contained a spiral pattern
  (fuelled by gas accretion/recycling and induced by gravitational
  perturbations to their disc).  Most S0-to-S transitions have likely already
  occurred, and this may have happened before the progenitor S0 galaxy became
  red/green and dust-poor, i.e., the fate of those which did not evolve.
  The S-to-S0 merger-induced transitions are expected to occur
  in the past, present, and future.  The green side of the `red sequence', or
  `green range', is due to gas-rich mergers building dust-rich S0 galaxies.
  These galaxies are an intermediate step on the path to merger-built E
  galaxies.  The `fading' trajectory shown in the figure applies to the past
  and future migration of blue S galaxies (some stellar mass loss from stellar
  winds is accounted for).  Although not shown, the slope of the red-sequence
  declines and possibly flattens at masses below $10^7$--$10^8$ M$_\odot$
  \citep[e.g.,][]{1968ApJ...151..105D, 2017ApJ...836..120R}.  Also not shown
  is the effect of tidal stripping, which shifts galaxies to the right and
  potentially upwards.  This schematic can be applied to understanding the
  CMD (Fig.~\ref{Fig2b}) and the CM$_{\rm bh}$D (Fig.~\ref{Fig6}).}
\label{Fig4}
\end{center}
\end{figure}

The lynchpin of the Tuning Fork diagram, connecting S and E galaxies, is the
(once thought to be singular in origin) S0 galaxy \citep{1925MNRAS..85.1014R,
  1936rene.book.....H}.  Astronomers used the S0 galaxy to try and discern how/if
galaxies may change from one type to another. 
\citet{1951ApJ...113..413S} envisaged that high-speed collision
between two spiral galaxies in a cluster may lead to the passage of the stars past
each other while the much higher number density of the gas particles
results in them colliding and being left behind.  They suggested this as a
means to produce gas-poor S0 galaxies.  This scenario was proferred before hot
X-ray gas halos in clusters were known.  \citet{1972ApJ...176....1G} and
\citet{1973MNRAS.165..231D} subsequently advocated for the more efficient
process in which the pervasive hot gas `ram-pressure' strips the cool
star-forming gas from S galaxies, and the mechanism proposed by
\citet{1951ApJ...113..413S} declined in popularity as the idea of altering
single S galaxies, rather than pairs, rose in popularity. It was also speculated that S
galaxies may collide to produce E-like galaxies
\citep[e.g.,][]{1954ApJ...119..206B, 1954Natur.173..818B,
  1954ApJ...119..232M}.  In group and field environments, where galaxy approach
speeds are sufficiently slow, a collision will result in the bulk of the
`over-shooting' stars falling
back in, resulting in a merger product that is gas- and dust-rich
\citep[e.g.,][]{1991AJ....101.2034M, 1995AJ....109..960W}.

Following these two popular mechanisms proposed for creating S0 galaxies,
\citet{2014MNRAS.440..889S} reported on these two suspected populations in the
`green valley': gas-depleted S
galaxies\footnote{As \citet{2010ApJ...721..193P} and others note, the gas may
be slowly consumed via dwindling star formation or rapidly removed from the
galaxy via various mechanisms.} and S galaxy merger products. 
\citet{2021MNRAS.508..895D} and others supported this duopoly. 
Departing from the notion that S0 galaxies are only a bridging population formed from
either merged or gas-stripped/starved S galaxies, the `Triangal' 
recognises a third type of S0 galaxy. 
It considers the bulk of the ($z=0$) dust-poor S0 galaxies to be 
faded S0 (not faded S) galaxies that are predominantly (by stellar mass)
primordial.

For dwarf S0 galaxies embedded in a cluster, they tend to be ram-pressure
stripped of their gas \citep{2003ApJ...591..167C}, dust-poor
\citep{2012MNRAS.421.3027B, 2015MNRAS.451.3815A}, and red/green. 
This does not rule out `downsizing' \citep[e.g.,][]{2000MNRAS.311..279M} 
or rejuvenation in other environments. 
If dwarf S0 galaxies are isolated, or still in the periphery of a cluster, 
they may still have an H{\footnotesize I} gas 
reservoir and low levels of current or recent star formation
\citep{2013A&A...552A...8D, 2017AJ....154...58H}. 
The gradual accrual of satellite galaxies and minor mergers will bring new stars and
gas and disturb high-angular momentum gas around the central galaxy. 
The starburst galaxy NGC~3034
\citep[M82:][not in sample]{2007PASP..119....1M}, with its nascent spiral arms
\citep{2005ApJ...628L..33M}, may be an example of such a transition galaxy.
If the galaxies are
in small groups, then this gas may form H{\footnotesize I} bridges with neighbouring
galaxies or robbed of some of its angular momentum through gravitational
interactions (with passing galaxies), leading it to fall onto its central
galaxy.  The extent of any ensuing star formation  will dictate how blue, and
thus how far into the `blue cloud', the galaxy evolves (Fig.~\ref{Fig4}).

The `Triangal' also builds on the phenomenon of S galaxy collisions
as a pathway to dust-rich S0 galaxies
\citep[e.g.,][]{2020MNRAS.498.2372D, 2021MNRAS.508..895D,
  2022MNRAS.515..201C}, rather than immediately building E galaxies
\citep{1990Natur.344..417W}, which require further mergers
to reduce/cancel the net angular momentum of the stars and erase the disc(s)
in the progenitor galaxies \citep{2003ApJ...597..893N}.
The discs of the merger-built S0 galaxies may help preserve the dust clouds, keeping them in orbit over
long durations, at least until their dust/gas is removed by, for example,
immersion in an X-ray hot halo of gas. 
The accumulation of dust-rich S0 galaxies at predominantly the {\it high-mass} end
of the S galaxy distribution (Fig.~\ref{Fig2b}) may reflect the rarity (but not absence)
of {\it low-mass} S galaxy mergers, expected to yield
low-mass dust-rich S0 galaxies.  This situation may arise in the Virgo Cluster 
because the high `velocity dispersion' of the galaxies leads to
fly-bys rather than collisions for the lower-mass galaxies with less
gravitational attraction.  Cold gas and dust removal may also be more
effective at lower galaxy masses. 
These evolutionary pathways have been shown schematically in Fig.~\ref{Fig3} 
and \ref{Fig4},
which builds on \citet[][their figure~10]{2007ApJ...665..265F} in several ways.

The CMD has long displayed a bimodal distribution
\citep[e.g.,][]{1964AJ.....69..635C, 2001AJ....122.1861S, 2007ApJ...665..265F}. 
Correcting for dust within the external galaxies,
as done here, a bimodal distribution in the CMD was also reported
by \citet{2009ApJ...699..105C}, and this was also seen using infrared
colours \citep{2014ApJ...794L..13A}
from the Wide Field Infrared Survey Explorer \citep[WISE:][]{2010AJ....140.1868W}.
The bimodal distribution is also apparent in Fig.~\ref{Fig2b} and
Fig.~\ref{Fig6}. 
However, one significant difference 
here is the notion that spiral galaxies start as spiral-less S0
galaxies.\footnote{Of course, when they first formed stars, the primordial S0 galaxies would
have been blue, and as such, the schema in Fig.~\ref{Fig4} does not quite capture that
element of the story.} 
The evolutionary chain commenced with what are now the old, metal-poor and dust-poor S0
galaxies in the local Universe. 
In the younger Universe, these disc galaxies  may have either been stripped of their (ability to acquire) gas --- and
thus effectively experienced a cessation of star formation --- or they may have grown
through acquisitions, becoming S and more massive S0 galaxies.  
When these galaxies reside within a hot
X-ray-emitting gas cloud, prevalent in galaxy cluster environments,
it tends to destroy the dust \citep{1979ApJ...231...77D} and
prevents gas from cooling to form new stars.  
With the bulk of their stars old and the galaxy having not
experienced major mergers, today's dust-poor S0 galaxies need never have been S
galaxies and are regarded as primordial (ancients, elders) left over from the young Universe.
As noted before, they need not all be old, with downsizing and delayed
creation capable of producing young(er) discs and even first-generation
galaxies from `dark galaxies'  
(e.g., O'Neil et al.\ 2024, in preparation; Soria et al.\ 2024, in preparation).

Another inclusion in Fig.~\ref{Fig4} is the explicit production of dust-rich S0 galaxies
from major wet mergers. 
Taken together with the above interpretation of the dust-poor S0 galaxies,
this likely contributes to, if not explains, the apparent population of both high- 
and low-mass S0 galaxies reported by \citet{1990ApJ...348...57V} and the
observations of ETG age reported by \citet{2001ApJ...553...90V}. 
Due in part to low levels of ongoing star formation
\citep{Graham-SFR}, the dusty S0 galaxies appear on the green side of the `red
sequence', referred to here as the `green range'.
Studying ETGs, \citet{1992AJ....104.1039S} revealed that 
S0 galaxies with signs of  merger-induced fine structure
tend to reside blueward of the `red sequence' in the CMD. 
There is also a tendency observed here for dust-rich ETGs to reside on the blue/green side of the `red
sequence'.
These dust-rich S0 galaxies are known merger remnants \citep{Graham-S0}. 
\citet{2019A&A...625A..94H} report that 
ETGs on the green side of the `red
sequence' often show evidence of dust extinction, which they take
as a signpost for young stellar populations causing the bluer colours.
Conceivably, some of the galaxies reported by \citet{2017MNRAS.466.2570B} 
to have central 
low ionisation emission-line regions (cLIERs) 
may not be spirals with bulges built via secular evolution
but lenticulars with merger-built bulges.

Among the sample of 88 ($=100-9-3$) Virgo Cluster ETGs
\citep{2006ApJS..164..334F}\footnote{Among the 16 to 21 Virgo Cluster ETGs
which do or may have dust, \citet{2006ApJS..164..334F} suggested that those
(2) with strong irregular dust lanes may evolve into those (4) with large
(kpc-sized) patchy discs of dust, before settling into those (3) with small
(few hundred parsecs), smooth, thin dust discs, i.e., those in the `n'
(nuclear) dust bin.} included here,
all four dIrr/dE transition objects (VCC~21, VCC~1512,
VCC~1499 and VCC~1779) may have weak dust features, and three of these four
have relatively blue $(g-i)_{\rm AB}$ colours for an ETG.  The exception is VCC~1512,
which is reported to have $(g-i)_{\rm AB} = 0.99$ mag, although
\citet{2016A&A...591A..38C} report 0.87 mag. 
The appearance of star-forming galaxies \citep{2014A&A...569A.124V} along the
green side of the `red sequence' can also be seen in \citet[][their
  figure~1]{2017ApJ...836..120R}.  While it is somewhat unclear which of
their 
star-forming galaxies might be S galaxies rather than dusty S0 galaxies, the absence of a
`blue cloud' --- in their sample of 404 galaxies \citep{2020ApJ...890..128F}
located around Virgo~A (aka M87) --- suggests few S galaxies are present.
In the SFR-mass diagram, 
the dust-rich S0 galaxies (e.g., NGC~1194, NGC~1316,
NGC~5018 and NGC~5128) also contribute toward the well-populated
`green mountain', located between the star-forming
main sequence and the true E galaxies 
\citep{2018MNRAS.473.3507E, 2018MNRAS.481.1183E, Graham-SFR}.
They are also seen here to shape a 
`green mountain' in the CMD after correcting their colour for dust, as
done for the LTGs. 

Differing from \citet{2014MNRAS.440..889S}, these merger-built S0 galaxies
tend to be located 
toward the top of the `blue cloud' of S galaxies. Rather than being a small
population that moves rapidly to the `red sequence', they appear to be a
substantial population that lingers at the green/red end of the `blue cloud',
modulo the adopted correction for dust in these galaxies. This is the opposite
behaviour to that concluded by \citet{2014MNRAS.440..889S}.
Furthermore, a sample of (potentially)
faded S galaxies are located on/near the `red sequence' rather than 
at the green/red end of 
the `blue cloud'. This, too, is the opposite finding that 
\citet{2014MNRAS.440..889S} reached, and it does not depend on a correction for dust in
these galaxies. 
Five dust-poor S0 galaxies (NGC: 1023; 3384; 4371; 4762; and 7332) had
previously stood out
by overlapping with the S galaxies in the $M_{\rm bh}$-$M_{\rm *,gal}$ diagram 
(Fig~\ref{Fig0}), 
making them good candidates for faded and transformed S galaxies.  {\it SDSS} colours
are available for all but NGC~1023, and they are highlighted in the
CMD (Fig.~\ref{Fig2b}).  Three are located at the high-mass end of the
dust-poor S0 galaxy sequence and have a red colour,
while the fourth (NGC~3384) is considered green. 
Most galaxies in the `green valley' sample of \citet{2014MNRAS.440..889S} had an
unknown morphological type, requiring assumptions about which 
were faded LTGs or merger-built ETGs, the two migratory scenarios that had been around
for decades. 
The present investigation advocates for a reversal of their conclusions as to
the S-to-S0 pathways through the `green valley'.

A few lower-mass dust-rich S0 galaxies trace a `green range'. 
This is not a discovery but perhaps a somewhat forgotten observation.
The existence of relatively blue E galaxies in the Millennium Galaxy Catalog (MGC) 
was called out by \citet{2007ApJ...657L..85D} and \citet{2009ApJ...699..105C}. 
Some of these blue dwarf galaxies \citep[e.g.,][]{2016MNRAS.457.1308M, 2019MNRAS.489.2830M}
may be dwarf S0 galaxies, for which the disc 
was inadvertently missed, rather than actual blue dwarf E galaxies.
Related are the blue compact dwarf (BCD) galaxies that are more knotty in structure but also 
possess an underlying backbone that is a disc \citep{2003ApJ...593..312C}. 

Stars in S galaxies eject metals, and before an S galaxy-collision-induced
burst of star formation in the merged system \citep{1992ApJ...400..153M}, the
gas clouds need to cool. This cooling would be associated with the gas-phase
metals condensing into dust particles, thus producing dust-obscured
starbursts \citep{1994ApJ...429..582C, 1996ApJ...464..641M,
  2006ApJS..163....1H}.  During a collision, the dramatic infall and cooling
of gas also produces central accretion discs around black holes and spurs the
quasar phenomenon in the disturbed remnant merger products
\citep[e.g.,][]{1989ApJ...336..681B, 1992ApJ...389..208B, 1998ApJ...505..159M,
  2012MNRAS.423...49K, 2023NatAs...7..463R} and, in some cases, an
ultraluminous infrared galaxy \citep[ULIRG:][]{1988ApJ...325...74S}. This may
also be the situation seen by \citet[][see their figure~10]{2016A&A...591A..38C}.
However, given that S galaxy major merger remnants still have substantial discs, not
all of the gas freefalls inward for consumption in a starburst.  Substantial
amounts of gas and dust remain in place due to their orbital angular momentum.
As the AGN fizzles out over time, the S0 merger remnant retains low levels of star
formation.  The subsequent merger of two or more of these S0 galaxies finally 
produces a massive E galaxy enshrouded in a hot gas halo that quenches star
formation and maintains the red galaxy colour. 
Low-level radio-mode (as opposed to gas-rich quasar-mode) heating from the
central BH, a so-called `Benson Burner'\footnote{\citet[][their footnote 33]{Graham:Sahu:22a}
introduced this term.}
 \citep{2003ApJ...599...38B, 2006MNRAS.370..645B}, 
appears capable of maintaining the hot X-ray gas halo, which keeps the galaxies quenched
and red by returning dust to metals in the gas phase and evaporating the cool
gas clouds \citep{2001ApJ...551..131C}. 

\citet{Graham-SFR} explored if the black hole mass or the $M_{\rm bh}/M_*$
ratio may dictate the SFR of galaxies through AGN feedback
\citep{2000MNRAS.311..576K, 2005Natur.433..604D, 2006MNRAS.365...11C,
  2009Natur.460..213C, 2012ARA&A..50..455F, 2014A&A...562A..21C}. 
It was found that it did not.  Instead, the 
SFR tracked the galaxy morphology more closely. Thus, the 
merger/accretion history of galaxies is more relevant than AGN feedback if
$M_{\rm bh}/M_*$ is a proxy for such feedback \citep{2016ApJ...830L..12T}.

\subsection{Future Work}

Future work could explore the evolution of 
the metallicity \citep[e.g.,][]{1992ApJ...398...69W, 2003MNRAS.339...63C, 2014PASA...31...36S},
and thus IMF \citep{2009A&A...504..373C} due to new generations of stars
coming, in part, from mergers.  
The low-metallicity quasi-primordial dust-poor S0 galaxies
\citep{2003AJ....125...66C, 2006AJ....132.2432L, 2013MSAIS..25...93S} 
would have a different IMF to the dust-rich S0 galaxies built from the merger of S galaxies.
A low metallicity 
would have resulted in the IMF having had a higher fraction of higher-mass 
stars than forms in S galaxies today \citep[e.g.,][]{2015ApJ...806L..31M, 2023Natur.613..460L}. 
The situation is compounded by the mass-metallicity and gas-dust trends observed in
disc galaxies today \citep[e.g.,][]{2008ApJ...678..804E, 2010MNRAS.408.2115M}. 
Along with the galaxies' internal evolution \citep{2014prpl.conf..149T}, this 
can impact the assigned $M/L$ ratios \citep[e.g.,][their figures~5--6]{2013MNRAS.430.2715I}. 
These different populations then find their 
way into E galaxies, which would have multiple metallicities. 
As galaxies evolve, one might expect the IMF to change with the $B/T$ ratio
and thus velocity dispersion, as observed by, for example,
\citet{2014MNRAS.438.1483S, 2020ARA&A..58..577S}.  
Following suggestions that a \citet{1955ApJ...121..161S} IMF is applicable to
ETGs, 
\citet[][their section~5.4]{2010MNRAS.404.2087B} address this IMF issue by
respectively adding 0.2 and 0.25 dex (in accord with a \citet{1955ApJ...121..161S} IMF)
to the masses of their S0 and E galaxies 
relative to the derivation obtained using a \citet{2003ApJ...586L.133C} IMF
that was maintained for the S galaxies. 
\citet[][their table~4]{2013MNRAS.430.2715I} provide zero-point calibrations
for the colour-mass relations appropriate for a \citet{1955ApJ...121..161S} IMF. 
Untangling samples of S0 galaxies with different creation histories should be an
important next step on this frontier, rather than considering all S0 galaxies
have, say, a \citet{1955ApJ...121..161S} or diet-Salpeter IMF.
IMF gradients and spaxel-by-spaxel variations may also abound
\citep[e.g.,][]{2018MNRAS.477.3954P, 2023arXiv231213355M}. 

For ETGs, it has been suggested that the gravitational potential energy, $\Phi
\propto M/R$, correlates better with colour than mass
\citep{2018ApJ...856...64B}, both of which are used in the CMD.  However, this
is problematic on several fronts. While the virial theorem implies $\sigma^2
\propto M/R$ for a pressure-supported system\footnote{The quantity $\sigma$ is
typically, albeit incorrectly, taken as the luminosity-weighted line of sight
velocity dispersion within $R$.}, S0 galaxies are predominantly
rotation-dominated systems, with disc-to-bulge mass ratios substantially
greater than 1. As such, the virial theorem does not apply, and using
$\sigma^2$ to trace the gravitational potential (and $\sigma^2/R_{\rm e,gal}$
to trace the surface mass density) is only meaningful for the E galaxies.
Moreover, even for the true E galaxies, changing $R$ from, say, the radius
enclosing 50 per cent of the light to some other percentage will
systematically, as a function of the E galaxy's S\'ersic index and thus
stellar mass \citep{1988MNRAS.232..239D, 1993MNRAS.265.1013C}, change the
estimated gravitational potential, surface mass density, and implied fraction
of dark matter \citep[][fig.~A1]{2019PASA...36...35G, 2023MNRAS.518.6293G}.
More meaningful measures of these quantities for the pressure-supported
element of ETGs will come from greater recognition of the multi-component
nature of galaxies.  Attempts to involve a third parameter (beyond morphology)
in the CMD is left for elsewhere.

For the past century, astronomers have frequently overlooked the presence of
discs in ordinary ETGs. 
This is also the case for some (most?) dwarf ETGs, which have been found to rotate
\citep[e.g.,][]{2002MNRAS.332L..59P, 2003AJ....126.1794G, 2009ApJ...707L..17T}.
Although dwarf S galaxies are uncommon, researchers have discovered weak
spiral patterns in some supposed dwarf ETGs \citep{2000A&A...358..845J,
  2002A&A...391..823B, 2003AJ....126.1787G},
 indicating the existence of discs within these galaxies. 
Connections in the CMD with dwarf
Irregular (dIrr) galaxies and blue compact dwarfs (BCDs) --- 
which also have a disc-like backbone and rotation \citep{2003ApJ...593..312C,
  2015A&A...577A..21C} --- 
and rotating low-mass blue/green ETGs \citep{2007ApJ...657L..85D,
  2009ApJ...699..105C, 2009AJ....138..579K, 2019MNRAS.489.2830M}, 
and dwarf spheroidal galaxies is deferred to elsewhere.
Nonetheless, for the low-mass (dwarf) galaxies, 
the role of accretion, mergers and fading, and concepts of
primordial versus evolved populations would benefit from further exploration.
For instance, are BCDs the product of disc$+$disc wet major mergers or
rejuvenated discs due to lesser accretion events
\citep[][and references therein]{2001ApJS..133..321C}? 
Discs can  largely be stripped 
away from galaxies, leaving behind somewhat naked bulges
\citep[aka `compact elliptical, cE:][]{2001ApJ...557L..39B}. 
or more violently threshed, leaving behind the tightly bound 
nuclear star cluster
\citep[NSC, aka `ultra-compact dwarf, UCD:'][]{2001ApJ...552L.105B}. 
Investigation of
galaxy-morphology-dependent $M_{\rm nsc}$-$M_{\rm bh}$ relations and 
the $M_{\rm ucd}$-$M_{\rm bh}$ relation \citep{2020MNRAS.492.3263G} 
will similarly be left for elsewhere. 
Fig.~\ref{Fig0} does not (yet) include dwarf galaxies. However, it 
suggests that disc-dominated dwarf S0 galaxies may be the place to search for
the largely missing population of intermediate-mass black holes
($10^2 < M_{\rm bh}/M_\odot < 10^5$). 

Future studies of the CMD may also benefit from introducing the wealth of
additional morphological information, such as spiral strength and bar
strength, as meticulously catalogued in the RC3 \citep{1991rc3..book.....D}
and elsewhere \citep[e.g.,][]{2015ApJS..217...32B}.
For instance, it is noted here that the 
dust-poor S0 galaxies tend not to be S0$^+$ galaxies but S0$^0$ or
S0$^-$,
while the dust-rich S0 galaxies can be S0$^-$, S0$^0$ and S0$^+$.

\section{Summary}
\label{Sec_Sum}

This paper paints morphologies into the CMD and colour-codes ETGs based on
their dust content.  Doing so has better revealed the substructure in the CMD.
Rather than just ETGs (no spiral pattern) versus
LTGs (contains a spiral pattern),
three types of S0 galaxy are considered:
dust-poor primordial S0 (not assumed to have once had a spiral pattern), 
dust-poor S0 (from faded S), 
and wet-merger built dust-rich S0 galaxies.
Progress with, and new detail in, the $M_{\rm bh}$-$M_{\rm *,sph}$ diagram
(Fig.~\ref{Fig0}) has aided adjustments of interpretation in the CMD,
including the notion of a buried `green mountain' of merger-built dust-rich S0
galaxies.

Within the CMD, the `blue cloud' would have been the dominant feature in the
early Gyrs of the Universe, with some galaxy discs forming spiral patterns
that survived until today and others neither able to sustain nor obtain
one. Those unable to have a spiral flourish will have fallen out of the `blue
cloud', but this does not mean they once possessed a spiral pattern. While
much work is focussed on the future evolution of today's S galaxies by trying
to shift them onto the `red sequence', less attention \citep[but
  see][]{2022MNRAS.511.5405B} seems to have been paid to the past evolution of
spiral-less disc galaxies.  Here, the evolutionary sequence known as the
`Triangal' (Fig.~\ref{Fig_schemas}) is mapped into the CMD (Fig.~\ref{Fig3}
and \ref{Fig4}), revealing how S galaxies can be a bridging population between
what are today's dust-poor and dust-rich S0 galaxies, as observed in the
$M_{\rm bh}$-$M_{*,sph}$ diagram (Fig.~\ref{Fig0}).

This paper introduces and checks the compatibility and applicability of a
revised schema which builds on the galaxy evolutionary tracks in the CMD presented by
\citet{2007ApJ...665..265F} and other works. It shows the Tuning Fork morphology
sequence (S - S0 - E) in the CMD 
\citep[e.g.,][]{2022A&A...666A.170Q} and expands on this to reveal the pathways captured by the 
`Triangal' \citep{Graham-triangal}.
This includes
S0 (now dust-poor) $\rightarrow$ S $\rightarrow$ S0 (dust-rich) $\rightarrow$
ES/E, encapsulating the
concept that spiral patterns likely emerged in spiral-less lenticular galaxies
due to internal instabilities or perturbations from giant molecular clouds or external
perturbations from gas accretion and minor mergers
\citep{1977ARA&A..15..437T}.  This pathway was addressed by 
\citet{2009AJ....138..579K} in terms of blue S0 galaxies at low masses. 
Fig.~\ref{Fig0} captures this and the major wet
merger-induced transition from S galaxy to dust-rich S0 galaxy
--- for which the turbulence in the new disc will be higher than in the S
galaxies and the velocity
dispersion elevated.  Past mergers of what are today's dust-poor S0 galaxies
(which were once star-forming galaxies) can also yield massive S0 galaxies,
and it has been suggested that this may explain compact massive
galaxies\footnote{Some `compact elliptical' (cE) dwarf galaxies may instead be
the bulges of heavily disc-stripped galaxies \citep{2001ApJ...557L..39B,
  2013pss6.book...91G}.}  \citep{Graham:Sahu:22b}. Additional mergers of
dust-rich S0 galaxies will further erode their discs to produce
spheroid-dominated ES and E galaxies, as seen in Fig.~\ref{Fig0}.
This provides a more nuanced scenario than envisioned by a statement like
`disc galaxy mergers produce E galaxies'. It also departs from the notion of
a single Hubble type for S0 galaxies and from de 
Vaucouleurs' T-types confining all S0 galaxies to a bridging population
between S and E galaxies, which is the main paradigm shift of the `Triangal'.

Here, the single `red sequence' is better regarded as two sequences: one for
(predominantly primordial) dust-poor S0 galaxies yielding the sloping
component of the `red sequence' and a `red plateau' defined by (pure) E
galaxies built via dry major mergers.  These E galaxies account for the
previously seen levelling off of colour in the distribution at high masses in
the CMD \citep[e.g.,][]{1968ApJ...151..105D, 2011MNRAS.417..785J}.  New in
regard to this is the clearer separation of galaxy types, facilitated by
careful multi-component decompositions of the galaxy light.

The appearance of the `Triangal' in the CMD, as shown here for the first time,
readily maps into the CM$_{\rm bh}$D. The `Triangal' serves as context 
for building on the apparent red and blue sequences in the $M_{\rm
  bh}$-$M_*$ diagram \citep{2016ApJ...817...21S} and the $M_{\rm bh}$-colour
diagram \citep{2020ApJ...898...83D}.  In
Section~\ref{Sec_col_bh}, (i) the designation of galaxy type in the $M_{\rm
  bh}$-colour diagram is expanded beyond simply LTG versus ETG and
dust corrections are applied to LTGs {\it and} ETGs 
with dust-rich discs.  This redefines the blue sequence in the $M_{\rm
  bh}$-colour diagram by incorporating the dust-rich S0 galaxies with the S galaxies, and it
redefines the `red sequence' in the $M_{\rm bh}$-colour diagram due to
galaxy-type-awareness, with a plateau evident for the (discless) E galaxies.
Finally, it also enables one to better understand the $M_{\rm bh}$-colour
relations in terms of galaxy evolution.

\section*{Acknowledgements}

The author is grateful for past conversations regarding galaxy colours with his initial PhD supervisor,
Natarajan (Vis) Visvanathan \citep{2002BAAS...34.1386F}. 
This work is based partly on observations made with the Spitzer Space
Telescope, which was operated by the Jet Propulsion Laboratory, California
Institute of Technology, under a contract with NASA.
This work has also used the NASA/IPAC Infrared Science Archive (IRSA)
and the NASA/IPAC Extragalactic Database (NED),
funded by NASA and operated by the California Institute of Technology.
This research has used the NASA/SAO Astrophysics Data System Bibliographic
Services.  It is also 
based partly on observations made with the NASA/ESA Hubble Space Telescope,
obtained from the Mikulski Archive for Space
Telescopes and the Hubble Legacy Archive.
Funding for SDSS-III has been provided by the Alfred P. Sloan Foundation, the
Participating Institutions, the National Science Foundation, and the
U.S. Department of Energy Office of Science.

\section{Data Availability}

The data used for this article are available in the references provided.
The S$^4$G dataset Digital Object Identifier (DOI) is 10.26131/IRSA425.

\bibliographystyle{mnras}
\bibliography{Paper-BH-col-mag}{}

\appendix

\section{Bulge-to-disc flux ratios}
\label{Sec_Apdx}

Use of the corrective formula (Eq.~\ref{Eq_inc})
for dust and inclination, collectively referred to as attenuation,
requires observed (pre-corrected) 
passband-specific $B/T$ flux ratios. The typical differences between these
ratios in various passbands are estimated here.

Due to the higher concentration of dust in the centres of galaxies
\citep[e.g.,][]{2013A&A...552A...8D, 2022MNRAS.513..389R}, the 
fractional correction for missing bulge light is greater than that for missing
disc light \citep[e.g.,][their figure~1]{2008ApJ...678L.101D}.
The young stellar population in the discs of
late-type S galaxies also acts to boost their disc flux and further lower their $B/T$
ratio at bluer wavelengths.  However, given that
late-type S galaxies (Sc-Sm) have small near-IR
$B/T$ ratios $<0.1$, the size of the attenuation correction
to the galaxy light remains rather insensitive to the $B/T$ ratio
in Eq.~\ref{Eq_inc}.  On the other hand, 
for massive S0 galaxies and early-type S galaxies (Sa-Sab), the
average near-IR $B/T$ ratio
can reach $\sim$0.2 to $\sim$0.3 \citep[figure~4 in][where
  $B/D \approx 1/4$ to 1/2]{Graham:Sahu:22a, 2008MNRAS.388.1708G}.  As such, the $B/T$
term in Eq.~\ref{Eq_inc} primarily affects these galaxies.

\begin{figure}
\begin{center}
\includegraphics[trim=0.0cm 0cm 0.0cm 0cm, width=1.0\columnwidth,
  angle=0]{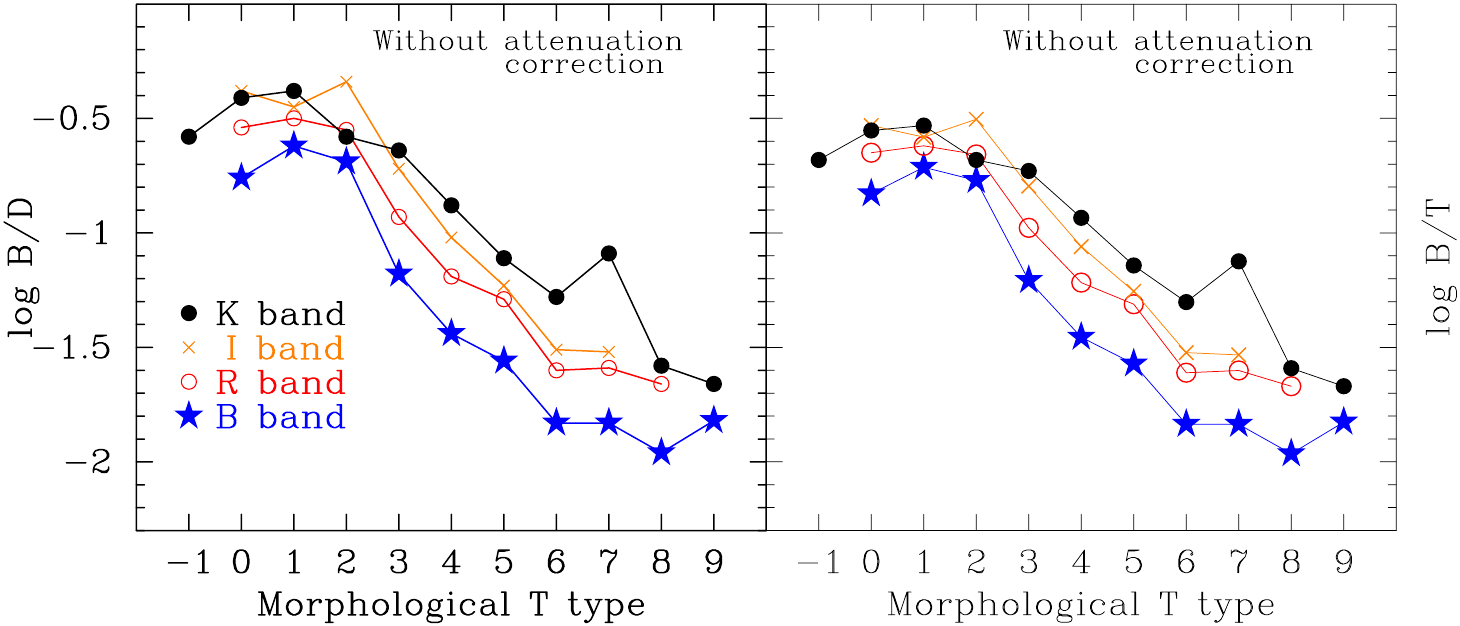} 
\caption{Median bulge-to-disc (left) and bulge-to-total (right)
  flux ratios in different passbands as a
  function of galaxy morphological T-type, such that the massive (likely
  merger-built and dust-rich) S0 galaxies have $T=-1$ \& 0, Sa=1, Sb=3,
  Sc=5, Sd=7, and Sm=9.
  Figure adapted from \citet[][their figure~7]{2008MNRAS.388.1708G} after removing
  the attenuation correction from \citet{2008ApJ...678L.101D} that accounted
  for dust and inclination.
}
\label{FigApp}
\end{center}
\end{figure}

Fig.~\ref{FigApp} displays the mean observed $B/T$ flux ratios of galaxies
before application of the attenuation correction from Eq.~\ref{Eq_inc}.
The data pertain to the samples presented 
in \citet{2008MNRAS.388.1708G}, consisting of 186, 193, 114, and 408 data
points in the $B$, $R$, $I$, and $K$-band, respectively.
The S0 galaxies are predominantly massive S0 galaxies with $B/T
\gtrsim$0.1--0.15. That is, they are not the low-mass dust-poor S0 galaxies.
Aside from the apparently deviant $I$ and $K$-band entries for the Sab (T=2)
galaxies, the difference in the $B/T$ ratios are smaller in the early-type
disc galaxies. This meshes well with the observation that the colours of
(the dust-free portions of) bulges and discs in early-type galaxies are not too
dissimilar from each other \citep{1996AJ....111.2238P}. 
For the late-type spirals, 
the $B/T$ ratios in the $I$, $R$, and $B$ bands are roughly 
0.7, 0.5 and 0.25 times the value in the $K$-band.
For the galaxies with big bulges (S0--Sab),
these numbers are taken to be 1.0, 0.8 and 0.5.

\bsp    
\label{lastpage}
\end{document}